\documentclass[12pt]{iopart}

\usepackage{graphicx}
\usepackage{fancyhdr}
\usepackage{amsfonts,amssymb}
\usepackage{color}

\begin{document}

\newcommand{\dR}{\mathbb R}
\newcommand{\dC}{\mathbb C}
\newcommand{\dZ}{\mathbb Z}
\newcommand{\id}{\mathbb I}
\newcommand{\dT}{\mathbb T}

\title[]{Evolution in bouncing quantum cosmology}

\author{Jakub Mielczarek$^{1,2}$ and W{\l}odzimierz Piechocki$^2$}

\address{$^1$Astronomical Observatory, Jagiellonian University, 30-244
Krak\'ow, Orla 171, Poland}

\address{$^2$Theoretical Physics Department, National
Centre for Nuclear Research,\\ Ho{\.z}a 69, 00-681 Warsaw,
Poland}

\date{\today}

\begin{abstract}
We present the method of describing an evolution in quantum
cosmology in the framework of the reduced phase space quantization
of loop cosmology. We apply our method  to the flat
Friedman-Robertson-Walker model coupled to a massless scalar
field.  We identify the physical quantum Hamiltonian that  is
positive-definite and generates globally an unitary evolution of
considered quantum system. We examine properties  of expectation
values of physical observables in the process of the quantum big
bounce transition. The dispersion of  evolved observables are
studied for the Gaussian state. Calculated relative fluctuations
enable an examination of  the semi-classicality conditions and
possible occurrence of the cosmic forgetfulness.  Preliminary
estimations based on the cosmological data suggest that there was
no cosmic amnesia. Presented results are analytical, and numerical
computations are only used for the visualization purposes. Our
method may be generalized to sophisticated cosmological models
including the Bianchi type universes.
\end{abstract}

\pacs{98.80.Qc,04.60.Pp,04.20.Jb}

\maketitle

\section{Introduction}

The loop quantum cosmology (LQC) method seems to be an efficient
method of quantization of cosmological models of general
relativity developed recently. Presently, we have two versions of
this method: standard LQC (see, e.g.
\cite{Ashtekar:2003hd,Bojowald:2006da,Ashtekar:2006wn} and
references therein) and nonstandard LQC
\cite{Dzierzak:2008dy,Dzierzak:2009ip,Malkiewicz:2009qv,Malkiewicz:2009zd,
Mielczarek:2010rq,Mielczarek:2010wu,Dzierzak:2009dj,Malkiewicz:2010py}.
For an extended {\it motivation} for developing the nonstandard
LQC we recommend an appendix of our paper
\cite{Malkiewicz:2009qv}.

The {\it standard} LQC has been developed by several groups around
the world in the last decade. This approach follows the Dirac program
in which one first identifies the kinematical Hilbert space ignoring the
dynamical constraints. Next, the dynamical constraint (choosing suitable
gauges leads to a single constraint) of the theory is promoted
into an operator acting in this space. Kernel of this operator is
used to construct the physical Hilbert space. Similarly,
observables are analyzed firstly on the kinematical phase space
and  suitable unitary transformation is constructed to map the
kinematical results into the physical ones.

The {\it nonstandard} LQC has been proposed recently. In the first
stage of this approach one prepares the classical formalism for
quantization: (1) Solutions to the Hamilton equations which
satisfy the dynamical constraints are found. In the case of
complicated system of equations, one may examine the structure of
the constraint surface by the phase portrait methods for dynamical
systems \cite{perko}. (2) Elementary Dirac observables on the
constraint surface are determined, which define the physical phase
space (PFS). (3) Physical observables are defined, i.e. the
observables which after quantization can be used for predicting
outcomes to be compared with observational data. Physical
observables are introduced as functions of elementary Dirac
observables and an evolution parameter.  In the second stage, one
quantizes the classical system: (1) Self-adjoint representations
of physical observables are constructed by using the
representation of the algebra of elementary observables. (2)
Eigenvalue problems for physical observables are solved to get
their spectra. (3) Evolution of expectation values of physical
observables is examined. It consists in finding the new
Hamiltonian of the classical theory that generates dynamics on the
physical phase space. This new Hamiltonian is no longer a
dynamical constraint of the classical theory. One finds a
self-adjoint representation of this Hamiltonian. Quantum
Hamiltonian is used, via Stone's theorem, to define an unitary
operator. This operator is used to examine an evolution of the
quantum system.

Preliminary results obtained in
\cite{Malkiewicz:2009qv,Malkiewicz:2010py} indicate that proposed
method of describing an evolution of a quantum cosmological system
is reasonable. In this paper we give {\it complete} presentation
of the evolution. We demonstrate that it is able to reveal
the details concerning the nature of the quantum big bounce
transition. In particular, we analyze quantum {\it fluctuations}
of physical observables in the propagation across the quantum
bounce from the past to the future time infinities.

In the standard LQC an evolution of cosmological system is
described quite differently. The kernel equation for the operator
constraint is used to construct, after some formal rearrangements,
an equation interpreted as  an evolution equation.  Unfortunately,
so obtained equation is usually so complicated that one can only
solve it by combined analytical and numerical methods.  This
is why the preliminary examination of the evolution of the
quantum FRW model has shown that classical big bang turns into
quantum big bounce transition \cite{Ashtekar:2006wn}, but could
not say anything specific about the nature of the quantum bounce.
In particular, an evolution of the dispersion effects of quantum
observables was not done satisfactory. Replacing an exact
Hamiltonian constraint of the FRW model by a simpler one
have enabled making some analytical analysis. This simplified
method for describing an evolution, called sLQC
\cite{Ashtekar:2007em}, has shown that the cosmic {\it amnesia}
 for the case of {\it semiclassical} states,
discovered earlier in analyzes of a simple cosmological toy model
\cite{BojowaldNature}, does not occur \cite{Corichi:2007am}.

Our nonstandard LQC method enables analytical studies, with an
exact expression for dynamical constraints, of subtle quantum
effects of specific cosmological model of  the universe. In this
paper, we consider the flat FRW model with a free massless scalar
field. The choice of the model results from the fact that the
dispersion effects  have been examined so far  mainly for this
model. On the other hand,  the model is quite simple and the FRW
symmetry is supported by the current observational cosmology.

In order to have our paper self-contained,  we recall in Sec. II A
some aspects of  derivation of the Hamiltonian {\it constraint} of
our nonstandard LQC \cite{Mielczarek:2010rq}. In Sec. II B we
introduce the notion of the {\it physical} Hamiltonian.  In Sec.
III we construct the physical {\it quantum} Hamiltonian and
examine its spectral properties. The {\it evolution} of quantum
FRW model is presented in Sec. IV, where we consider the {\it
dispersion} of physical observables. We also briefly evoke the
problem of time. The relative {\it fluctuations} of observables,
for the Gaussian state, are considered as a function of time in
Sec. V. We conclude in the last section.  Appendix A includes the
derivation of the formulas used in the  section on quantum
fluctuations.

\section{Classical dynamics}

\subsection{Hamiltonian constraint}

The gravitational part of the classical Hamiltonian, in the
Ashtekar variables $(A^i_a,E^a_i)$, is the sum of the first class
constraints
\begin{equation}\label{conham}
H_{g} = \frac{1}{16\pi G} \int_{\Sigma} d^3{\bf x}
(N^iC_i+N^aC_a+NC),
\end{equation}
where $\Sigma$ is the space-like part of spacetime $\dR \times
\Sigma$, and where $C_i$ and $C_a$ denote the Gauss and the
spatial diffeomorphisms constraints, respectively. For considered
FRW model gauges are chosen in such a way that  $C_i$ and  $C_a$
constraints are automatically fulfilled. The only nontrivial part
is the scalar constraint $C$ so the Hamiltonian reads
\begin{equation}
H_{g} = -\frac{1}{\gamma^2}\frac{1}{16\pi G} \int_{\Sigma} d^3{\bf
x}\frac{1}{\sqrt{|\det E|}} E^a_i E^b_j \epsilon^{ij}_k F^k_{ab},
\label{Ham0}
\end{equation}
where $\gamma$ is the Barbero-Immirzi parameter, and where
$F^i_{ab}$ is the curvature of  $SU(2)$ connection $A^i_a$.

In LQC the gravitational degrees of freedom are parametrised by
holonomies $h_i$ and fluxes $F_i$, which are functionals of the
Ashtekar variables. These non-local functions are used to
construct a non-perturbative  theory. The holonomies and fluxes
are the $SU(2)$ variables satisfying the holonomy-flux algebra. In
the highly symmetric spaces, like the FRW model considered here,
the forms of these functions are simple and known.

In particular, in the flat FRW model the flux may be parametrised
by the  $v$ variable and the holonomy is expressed in terms of the
$\beta$ variable \cite{Malkiewicz:2009qv}. The variable  $v$  is a
physical {\it volume} defined as follows
\begin{eqnarray}\label{v1}
    v := \int_\mathcal{V} dx_1 dx_2 dx_3 \sqrt {det \,q_{ab}} =
    a^3 \int_\mathcal{V} dx_1 dx_2 dx_3 \sqrt {det \,q_{ab}^0}
    =: a^3\, V_0,
\end{eqnarray}
where $\mathcal{V}\subset \Sigma$ is an elementary {\it cell} in
the space with topology $\dR^3$; $(x_a) = (x^a) = (x^1, x^2, x^3)$
are Cartesian coordinates; $q_{ab}:= a^2 \,q_{ab}^0$ is a physical
3-metric; $a$ is a scale factor; $q_{ab}^0 dx^a dx^b:= dx_1^2 +
dx_2^2 + dx_3^2$ defines a fiducial 3-metric; $V_0$ is a fiducial
volume (it does not occur in final results). The $\beta$ variable,
in the limit $\beta \rightarrow 0$, is linked to the Hubble factor
$H =\dot{a}/a$ via the relation $\beta = \gamma H$.

In order to express the Hamiltonian, Eq. (\ref{Ham0}), in terms of
holonomies and fluxes, the procedure of regularization has to be
applied, which introduces a new scale to the theory, namely the
parameter $\lambda$. This can be understood as the length scale of
the lattice discretization. The applied procedure of
regularization and rewriting the Hamiltonian in terms of
holonomies and fluxes is the same as known from the standard LQC
(see, Appendix A of \cite{Mielczarek:2010rq}).
 The obtained gravitational
part of the Hamiltonian reads \cite{Ashtekar:2006wn}
\begin{eqnarray}
H^{(\lambda)}_g =  -\frac{vN}{32\pi^2 G^2\gamma^3\lambda^3}
\sum_{ijk} \epsilon^{ijk} \rm{tr} \left[h_{\Box_{ij}}h_k \left\{
(h_k)^{-1} ,v\right\}\right], \nonumber  \\ \label{hgl}
\end{eqnarray}
where $h_{\Box_{ij}} = h_i h_j (h_i)^{-1} (h_j)^{-1}$ is the
holonomy around the square loop $\Box_{ij}$ (for more details see
\cite{Dzierzak:2009ip}), and $N$ is the lapse function.  The
elementary holonomy in the i-th direction reads
\begin{equation}
h_i = \cos\left(\frac{\lambda
\beta}{2}\right)\mathbb{I}+2\sin\left( \frac{\lambda
\beta}{2}\right) \tau_i \label{hol}
\end{equation}
where $\tau_i=-\frac{i}{2} \sigma_i$ ($\sigma_i$ are the Pauli
matrices). The holonomy (\ref{hol}) is calculated in the
fundamental representation of $SU(2)$. The factor $\lambda$ is the
parameter of the theory that may be related with the minimum area
of the loop.  It is expected that $\lambda \sim l_{\rm{Pl}}$,
but its precise value has to be fixed observationally.

In the model considered in this paper, the total Hamiltonian is
the sum of the gravity $H^{(\lambda)}_g$ and  free scalar field
$H_{\phi}:= N p^2_{\phi}/(2 v)$ terms. The insertion of the
elementary holonomy (\ref{hol}) into Eq. (\ref{hgl}) leads (for
details, see Appendix A of \cite{Mielczarek:2010rq}) to the
expression\footnote{In the rest of the paper we choose
$\;c=1=\hslash\;$ and $\;G = l^2_{\rm{Pl}}=1/m^2_{\rm{Pl}}\;$
except where otherwise noted.}:
\begin{equation}\label{con1}
H^{(\lambda)} = N\left( - \frac{3}{8\pi l_{\rm{Pl}}^2 \gamma^2}
\frac{\sin^2 (\lambda\beta)}{\lambda^2}v+\frac{p^2_{\phi}}{2v}
\right) \approx 0,
\end{equation}
where the sign ``$\,\approx\,$'' reminds that the Hamiltonian is a
{\it constraint} of considered gravitational system.

The Hamilton equation takes the following form\footnote{$f$ is a function on phase
space.}
\begin{equation}
\label{HamiltonEq}
\frac{df}{dt}=\{f,H^{(\lambda)}\},
\end{equation}
where
\begin{equation}\label{nawiasP}
\{ \cdot , \cdot \} := 4\pi G \gamma \left[\frac{\partial \cdot }
{\partial \beta} \frac{\partial \cdot }{\partial v}-
\frac{\partial \cdot }{\partial v}\frac{\partial \cdot }{\partial
\beta} \right]
 + \frac{\partial \cdot }{\partial \phi} \frac{\partial \cdot }
 {\partial p_{\phi}}-
\frac{\partial \cdot }{\partial p_{\phi}}\frac{\partial \cdot }
{\partial \phi}.
\end{equation}
For the lapse function $N=1$ in Eq. (\ref{con1}), the time $t$ in
equation (\ref{HamiltonEq})  is the \emph{coordinate} time. Using
Eqs. (\ref{HamiltonEq}) and (\ref{v1}) we get the expression for
the Hubble factor, $H$, as follows
\begin{equation}
H:= \frac{\dot{a}}{a} = \frac{1}{3v}\frac{dv}{dt} =
\frac{1}{\gamma} \frac{\sin(2\lambda \beta)}{2\lambda}.
\label{Hubbledef}
\end{equation}
We have the expansion period ($H>0$) for $\beta \in
(0,\frac{\pi}{2\lambda})$, and the contraction ($H<0$) for $\beta
\in(\frac{\pi}{2\lambda},\frac{\pi}{\lambda})$. The
present classical phase of expansion corresponds to the limit
$\beta \rightarrow 0$. In this limit, due to (\ref{Hubbledef}), we
get $H=\beta/\gamma$. This relation can be also obtained
from (\ref{Hubbledef}) by shrinking the regularization parameter
$\lambda$ to zero.

The coordinate time $t$ used in cosmological observations and the
{\it intrinsic} time (to be introduced  later) preferred in
theoretical considerations  will be related via
\begin{equation}
\frac{d\phi}{dt} = \{\phi,H^{(\lambda)}\} = \frac{p_{\phi}}{v}.
\label{dphidt}
\end{equation}
Finding an explicit formula for $v$ in terms of $\phi$ (see next
subsection) gives more insight into this relationship.

\subsection{Physical Hamiltonian}

The {\it kinematical} phase space $\mathcal{F}_k$ of the system
can be parametrized by four independent variables $\beta, v,
\phi,$ and $ p_{\phi}$. If these variables satisfy the constraint
(\ref{con1}), they can be used to parametrize  the {\it physical}
phase space $\mathcal{F}_p$ which is thus three dimensional. In
the {\it relative} dynamics one canonical variable is used to
parametrize all others.  Choosing $\phi$ to play such a role
enables an integration of the system and finding the elementary
Dirac observables by the method presented in
\cite{Dzierzak:2009ip}:

\begin{eqnarray}
\mathcal{O}_1 &=& p_{\phi},\label{obs1} \\
\mathcal{O}_2 &=& \phi-\frac{\rm{sgn}(p_{\phi})}{\sqrt{12 \pi} l_{\rm{Pl}}}
\rm{arth} \left(  \cos \left( \beta\lambda \right) \right), \label{obs2} \\
\mathcal{O}_3 &=&\rm{sgn}(p_{\phi}) \ v \frac{\sin(\lambda
\beta)}{\lambda},\label{obs3}
\end{eqnarray}
which can be used to parametrize the phase space of the relative
dynamics. Equations (\ref{obs1})-(\ref{obs3}) take into account
the constraint (\ref{con1}) so we have the relation
\cite{Dzierzak:2009ip}
\begin{equation}\label{con2}
\mathcal{O}_3 = \kappa \gamma \mathcal{O}_1,
\end{equation}
where $\kappa^2 := 4 \pi G/3$.   Thus, the physical phase space of
the relative dynamics is two dimensional. It can be parametrized
by two independent elementary observables, for instance
$\mathcal{O}_2$ and $\mathcal{O}_3$.

The observables are {\it constants} of motions since, by the
definition of Dirac's observables, we have
\begin{equation}\label{conM}
\dot{\mathcal{{O}}}_k = \{\mathcal{O}_k,H^{(\lambda)}\} = 0,~~~~~~k
= 1,2,3 .
\end{equation}
Thus, acting with $d/d \phi $ on Egs. (\ref{obs1})-(\ref{obs3})
leads to
\begin{eqnarray}
0 &=& 1 +\frac{\rm{sgn}(p_{\phi})}{\sqrt{12 \pi} l_{\rm{Pl}}}
\frac{\lambda}{\sin(\lambda\beta)}
\frac{d\beta}{d\phi},\label{e1} \\
0 &=& \lambda \cos(\lambda\beta)\frac{d\beta}{d\phi} v +
\sin(\lambda\beta)\frac{dv}{d\phi}.\label{e2}
\end{eqnarray}
We wish to emphasize again that (\ref{e1}) and (\ref{e2}) do include the
constraint (\ref{con1}).  In redefined variables
\begin{eqnarray}\label{redpar}
Q := \beta,~~~~ P :=\frac{1}{4\pi l_{\rm{Pl}}^2 \gamma} v,~~~~ T
:= - \rm{sgn}(p_{\phi}) \sqrt{3 \pi } l_{\rm{Pl}}^2 \phi,
\end{eqnarray}
equations (\ref{e1}) and (\ref{e2}) can be rewritten
formally in the form
\begin{eqnarray}
\frac{dP}{dT} &=& - \frac{\partial H_{\lambda}}{\partial Q}, \label{Pv} \\
\frac{dQ}{dT} &=& \frac{\partial H_{\lambda}}{\partial P},\label{Qv}
\end{eqnarray}
where
\begin{equation}\label{cH}
H_{\lambda} :=\frac{2}{\lambda\sqrt{G}} P \sin(\lambda Q),
\end{equation}
plays the role of the Hamiltonian of the system {\it devoid} of
the dynamical constraint\footnote{Proving the equivalence of Eqs.
(\ref{Pv})-(\ref{Qv}) with Eqs. (\ref{e1})-(\ref{e2}) needs using
combination of (\ref{e1}) and (\ref{e2}).}.

This way we have turned the system {\it with} constraint into the
system {\it without} constraint. One can say that we have obtained
the system in which the dynamical constraint has been {\it
solved}.

Let us find the solution to the system defined by (\ref{Pv})-(\ref{cH}).
By direct integrations, we find
\begin{equation}\label{vQ}
Q = \frac{2}{\lambda} \arctan \exp \Big(
\frac{2}{\sqrt{G}}(T-T_0)\Big).
\end{equation}
Similarly, we get
\begin{equation}\label{vP}
P = P_0 \cosh \Big( \frac{2}{\sqrt{G}}(T-T_0)\Big),
\end{equation}
where $P_0$ and $T_0$ are constants of integration.

The parameter $T$ is the intrinsic time for the relative
dynamics under considerations.  Using(\ref{dphidt}),
(\ref{redpar}) and  (\ref{vP}) we relate this time with the
coordinate time $t$:
\begin{equation}\label{trel}
t=t_0 - 2 \sqrt{\frac{\pi G}{3}} \frac{\gamma
P_0}{|p_{\phi}|}\sinh \Big( \frac{2}{\sqrt{G}}(T-T_0)\Big).
\end{equation}
Since $P_0 > 0$ (as $P$ corresponds to the volume variable) and
the r.h.s. of Eq. (\ref{trel}) monotonically decreasing with $T$,
one can see that the directions of $t$ and $T$ are quite the
opposite. Thus, the phase of expansion in coordinate time $t$
corresponds to the contraction phase in intrinsic time $T$ and
vice versa.  One can give the following interpretation: we define
the Hubble parameter, $h$, in terms of the intrinsic time $T$, in
analogy to (\ref{Hubbledef}), to get\footnote{Now, we use the
Hamiltonian (\ref{cH}).}
\begin{equation}
h:=\frac{1}{3v} \frac{dv}{dT} = -\frac{2}{3} m_{\rm{Pl}}
\cos(\lambda Q). \label{Hubbleh}
\end{equation}
In contrast to the coordinate time case (\ref{Hubbledef}), we have
a contraction ($h < 0$) for $\;\lambda Q \in (0,\pi/2),\;$ and an
expansion ($h > 0 $)  for $\;\lambda Q \in (\pi/2, \pi)\;$. Using
this we can say again that the directions of the intrinsic time
$T$ and the coordinate time $t$ are the {\it opposite}.

It is commonly known that  Hamiltonian of a physical physical
system (being a part of some larger system) should be bounded from
below, otherwise it would be dynamically {\it unstable} as the
lowest energy state would not exist. It took much effort to prove
that asymptotically flat spacetimes  may have this property (see,
e.g. the positive energy theorem \cite{posen}). In the case of
cosmology, the situation is quite different since the system,
being the entire universe, is not a part of some bigger system. In
what follows we assume that our model of the universe is an
isolated system. Since we consider a model which is a Hamiltonian
system, its total energy must be conserved. Such a system (with
fixed spacetime geometry and topology) cannot make transition to a
state with lower or higher energy.  The only reasonable
requirement is that it should have finite value.  However, our
{\it classical} Hamiltonian, defined by (21), is positive-definite
since $\lambda Q\in (0,\pi)$ and $P> 0$, which enables  reasonable
interpretation of the model at the classical level. The values of $Q=0$
and  $Q=\pi/\lambda$ can be approached by the classical trajectories
only asymptotically. Owing to this, we {\it postulate} that the corresponding
{\it quantum} Hamiltonian should be positive-definite too. At this stage
we introduce the notion of the {\it physical}  Hamiltonian. It is defined to be a
positive-definite Hamiltonian that {\it generates} dynamics of the
system. The property of being positive-definite is specific to our
cosmological model and is devoid of basic importance.

The positivity of the Hamiltonian (\ref{cH}) was achieved by
introducing an intrinsic time that has an opposite sine to the
metric time.  It is possible to redefine intrinsic time $T
\rightarrow -T$, such that the directions of $T$ and $t$ will be
the same. However, in such a case the Hamiltonian (\ref{cH}) would
be multiplied by minus one so  $H_{\lambda}$  no longer would be
positive-definite.

The above reasoning presents the key ideas underlying the reduced
phase space (RPS) approach of the classical level. In what follows
we present the RPS quantization.

\section{Quantum Hamiltonian}

In what follows we use the Hilbert space $\mathcal{H} =
L^2([0,\pi/\lambda],dQ)$ so it has the scalar product
\begin{equation}
\langle f | g \rangle := \int_0^{\pi/\lambda} \overline{f}  g dQ.
\end{equation}

The quantum  Hamiltonian corresponding to the classical one
(\ref{cH}) is defined in a {\it standard} way to be
\begin{equation}\label{fH}
\hat{H}_{\lambda} :=\frac{m_{\rm{Pl}}}{\lambda}\left(\hat{P}\;
\widehat{\sin(\lambda Q)} + \widehat{\sin(\lambda Q})\hat{P}
\right).
\end{equation}

The classical canonical variables $Q$ and $P$ satisfy the algebra
$\{Q,P\}=1$. Choosing the Schr\"{o}dinger representations for
these variables
\begin{equation}\label{sr}
\hat{Q} \phi (Q):= Q  \phi (Q),~~~~~\hat{P} \phi (Q):= -i
\frac{d}{dQ} \phi (Q),
\end{equation}
where $\phi \in \mathcal{H}$,  gives formally $[\hat{Q},\hat{P}] =
i \hat{\mathbb{I}}$.

An explicit form of the operator (\ref{fH}) is easily found to be
\begin{equation}\label{eH}
\hat{H}_{\lambda}\psi = -\frac{i}{\lambda \sqrt{G}}\left(
2\sin(\lambda Q) \frac{d}{dQ}+\lambda \cos(\lambda Q)
\right)\psi,
\end{equation}
where $\psi \in D \subset \mathcal{H}$, and where $D$ is some {\it
dense} subspace of $\mathcal{H}$. In what follows we wish to
determine $D$ which is the domain of {\it self-adjointness} of the
operator $\hat{H}_{\lambda}$.

\subsection{Eigenproblem for the Hamiltonian}

The eigenequation for the operator $\hat{H}_{\lambda}$ reads

\begin{equation}
-\frac{i}{\lambda \sqrt{G}}\left( 2\sin(\lambda Q) \frac{d\Psi
}{dQ}+\lambda \cos(\lambda Q)\Psi \right)  = E \Psi.
\end{equation}
The solution of the above equation is given by
\begin{equation}
\Psi_E = \Psi_0  \frac{1}{\sqrt{\sin(\lambda Q)}} \exp\left\{ \frac{1}{2} i \sqrt{G} E \ln
\left| \tan\left( \frac{\lambda Q}{2} \right) \right| \right\},
\end{equation}
where $\Psi_0$ is a constant of integration. Let us calculate
\begin{eqnarray}
\langle \Psi_E | \Psi_{E'} \rangle =
|\Psi_0|^2 \int_0^{\pi/\lambda} dQ \frac{\exp\left\{ \frac{1}{2} i \sqrt{G} (E'-E) \ln \left| \tan\left(
\frac{\lambda Q}{2} \right) \right| \right\}  }{\sin(\lambda Q)}.
\end{eqnarray}
Defining
\begin{equation}
x:=  \frac{\sqrt{G}}{2} \ln \left| \tan\left( \frac{\lambda Q}{2} \right) \right|
\end{equation}
once can rewrite the above integral into the form
\begin{eqnarray}
\langle \Psi_E | \Psi_{E'} \rangle &=& |\Psi_0|^2 \frac{2}{\lambda \sqrt{G}} \int_{-\infty}^{+\infty}\exp\left\{ i (E'-E) x \right\} dx  \nonumber \\
&=& |\Psi_0|^2 \frac{2}{\lambda \sqrt{G}} 2\pi \delta(E'-E).
\end{eqnarray}
By choosing
\begin{equation}
\Psi_0 = \sqrt{ \frac{\lambda \sqrt{G}}{4\pi} },
\end{equation}
we get
\begin{equation}
\Psi_E(Q) =\langle Q  |  \Psi_E \rangle =  \sqrt{ \frac{\lambda
\sqrt{G}}{4\pi \sin(\lambda Q)} } \exp\left\{ \frac{1}{2} i
\sqrt{G} E \ln \left| \tan\left( \frac{\lambda Q}{2} \right)
\right| \right\}.
\label{eigenstate}
\end{equation}
The states $\Psi_E$ given by equation (\ref{eigenstate}) satisfies
the orthonormality condition in the form
\begin{eqnarray}
\langle \Psi_E | \Psi_{E'} \rangle = \delta(E'-E).
\end{eqnarray}
Based on this, for the eigenstates, we obtain
\begin{equation}
\langle \Psi_E | \hat{H}_{\lambda} \Psi_{E'} \rangle -\langle \hat{H}_{\lambda}  \Psi_E
| \Psi_{E'} \rangle= (E'-E)\delta(E'-E) =0,
\end{equation}
which means that the Hamiltonian is {\it symmetric} on the space
of the eigenstates.

\subsection{Symmetricity}

We define the domain of $\hat{H}_{\lambda}$ as follows
\begin{equation}
\label{dom} D(\hat{H}_{\lambda}):= {\rm span} \{ \varphi_k,~~k
\in \mathbb{Z} \},
\end{equation}
where
\begin{equation}\label{sep}
    \varphi_k(Q):= \int_{-\infty}^\infty  c_k (E)
    \Psi_{E}(Q)\; dE,~~~c_k \in C^\infty_0 (\dR).
\end{equation}
It is clear that $D(\hat{H}_{\lambda})\subset
L^2([0,\pi/\lambda],dQ)=\mathcal{H}$ is a dense subspace of
$\mathcal{H}$, and an action of the unbounded operator
$\hat{H}_{\lambda}$ does not lead outside of
$D(\hat{H}_{\lambda})$.

For any $\varphi_k$ and $\varphi_l$  we have
\begin{eqnarray}
\langle \varphi_k | \hat{H}_\lambda \varphi_l \rangle -\langle
\hat{H}_\lambda \varphi_k | \varphi_l \rangle= \nonumber \\
=\int_{-\infty}^{+\infty}\int_{-\infty}^{+\infty} \overline{c_k(E)}
c_l(E') \underbrace{(E'-E)\delta(E-E')}_{=0} dE dE'=0.
\end{eqnarray}
Therefore, the operator $\hat{H}_{\lambda}$ is symmetric on $D(\hat{H}_{\lambda})$.

\subsection{Self-adjointness}

To examine the self-adjointness of the unbounded operator
$\hat{H}_\lambda$, we first identify the deficiency subspaces,
$\mathcal{K_\pm}$,  of this operator \cite{RaS}
\begin{equation}\label{self1}
\mathcal{K_\pm}:= \{g_\pm \in D(\hat{H}_\lambda^\ast)~|~\langle
g_\pm|(\hat{H}_\lambda \pm i \id)\varphi,\rangle =0, \;\forall
\varphi \in D(\hat{H}_\lambda\},
\end{equation}
where  $D(\hat{H}_\lambda^\ast):=\{f\in
L^2([0,\pi/\lambda],dQ)~|~\exists !f^{\ast}~\langle
f^{\ast}|g\rangle=\langle f|\hat{H}_{\lambda} g\rangle,~\forall
g\in D(\hat{H}_{\lambda})\}$, and where
\begin{equation}\label{posD}
f(Q):=  \int_{-\infty}^\infty  b (E)
    \Psi_{E}(Q)\; dE,~~~b \in C^\infty_0 (\dR).
\end{equation}

For each $ \varphi_k \in D(\hat{H}_{\lambda})$, defined by
(\ref{sep}), we have

\begin{eqnarray}
0 &=& \langle g_\pm|(\hat{H}_\lambda \pm i \id)\varphi_k \rangle
\nonumber
\\& = &  \int_{0}^{\pi/\lambda} dQ
\int_{-\infty}^{\infty}dE_1 \int_{-\infty}^{\infty}dE_2 \;b_\pm
^\ast (E_2)\Psi_{E_2}^\ast(Q)(\hat{H}_\lambda \pm i \id)c_k
(E_1)\Psi_{E_1}(Q) \nonumber
\\& = & \int_{-\infty}^{\infty}dE_1 \int_{-\infty}^{\infty}dE_2 \;b_\pm
^\ast (E_2)c_k (E_1)(E_1 \pm i )\int_{0}^{\pi/\lambda}
\Psi_{E_2}^\ast(Q) \Psi_{E_1}(Q) dQ \nonumber
\\&=& \int_{-\infty}^{\infty}dE (E \pm i )b_\pm
^\ast (E)c_k (E)~~~~\Longrightarrow~~~~~b_+ = 0 = b_-
\label{selfH}.
\end{eqnarray}
Thus, the deficiency indices $n_\pm := dim [\mathcal{K}_\pm]$ of
$\hat{H}_\lambda$ satisfy the relation: $n_+ = 0 = n_-$, which
proves that the operator $\hat{H}_\lambda$ is {\it essentially}
self-adjoint on $D(\hat{H}_\lambda)$. One can argue that the whole
spectrum of $\hat{H}_\lambda$ belongs to the real line.

\subsection{Physical Hamiltonian}

The {\it classical} physical   Hamiltonian $H_{\lambda}$ is
positive-definite. The corresponding self-adjoint operator
$\hat{H}_{\lambda}$ has however eigenvalues $E \in \dR $. We
therefore introduce the {\it quantum}  physical Hamiltonian
$\hat{\mathbb{H}}$ by requiring that it has only {\it positive}
eigenvalues. It is defined as follows
\begin{equation}
\hat{\mathbb{H}} |  \Psi_{E} \rangle := |E| |  \Psi_{E}
\rangle,~~~~E>0,
\end{equation}
where $|  \Psi_{E} \rangle $ is the eigenvector of the Hamiltonian
$\hat{H}$ corresponding to the eigenvalue $E$. It is clear that
the spectrum of the operator $\hat{\mathbb{H}}$ is doubly
degenerate. For any state $\Psi  =\int_{-\infty}^{+\infty} c(E)
\Psi_{E}\ dE   \in D(\hat{\mathbb{H}})= D(\hat{H}_{\lambda})$,
where $c \in C^\infty_0(\mathbb{R})$,  we get
\begin{equation}
\langle \Psi | \hat{\mathbb{H}} | \Psi \rangle =
\int_{-\infty}^{+\infty}dE |c(E)|^2 |E| > 0.
\end{equation}

\section{Evolution of quantum system}

Making use of the Stone theorem \cite{RaS}, we  define the unitary
operator of an {\it evolution} as follows
\begin{equation}\label{evo}
\hat{U}(s) := e^{- i s \hat{\mathbb{H}}},
\end{equation}
where $s \in \dR$ is a `time' parameter. The state at any moment
of time can be found as follows
\begin{equation}
| \Psi(s) \rangle = \hat{U}(s)| \Psi(0) \rangle = e^{- i s
\hat{\mathbb{H}}}| \Psi(0) \rangle .
\end{equation}
The {\it minus} sign in (\ref{evo}), of the exponential
function of an operator, is essential as only in this case an
infinitesimal version of $\hat{U}(s)$ acting on $| \Psi(s)
\rangle$ leads to the Schr\"{o}dinger equation.

Let us consider a superposition of the Hamiltonian eigenstates
\begin{equation}
| \Psi(0) \rangle = \int_{-\infty}^{+\infty} dE c(E) |
\Psi_{E} \rangle
\end{equation}
at $s=0$. Then evolution of this state is given by
\begin{eqnarray}\label{sup}
|\Psi(s)\rangle = \hat{U}(s) |\Psi(0)\rangle =\int_{-\infty}^{+\infty} dE
c(E) e^{- i s \hat{\mathbb{H}}} |\Psi_{E}\rangle \nonumber \\
=\int_{-\infty}^{+\infty}  dE c(E) e^{- i s
|E|}|\Psi_{E}\rangle.
\end{eqnarray}
Normalization of this state is given by the following condition
\begin{eqnarray}
\langle \Psi(s)|\Psi(s)\rangle =
\int_{-\infty}^{+\infty}\int_{-\infty}^{+\infty} dEdE' \overline{
c(E)} c(E')
 e^{- i s (|E|-|E'|)}\underbrace{\langle \Psi_{E} |\Psi_{E'}\rangle}_{=\delta(E'-E)}  \nonumber \\
 =\int_{-\infty}^{+\infty} dE | c(E)|^2 = 1.
\label{PsiNorm}
\end{eqnarray}
Now, let us assume that the superposition (\ref{sup}) has a form
of the Gaussian packet with the profile defined to
be\begin{equation}\label{gauss}
c(E)  := \left(  \frac{2 \alpha}{\pi} \right)^{1/4}
e^{-\alpha\;(E-E_0)^2},
\end{equation}
centered around $E_0$ with the dispersion  parametrized by
$\alpha$. \ The normalization factor $\left(  2 \alpha/\pi
\right)^{1/4} $ is due to the condition (\ref{PsiNorm}).  In the
rest of this paper we  study properties of this state only.
Assuming that $E_0\sqrt{\alpha} \gg 1$, one can approximate
\begin{eqnarray}
\Psi(Q,s) &=& \langle Q |\Psi(s)\rangle \left(  \frac{2 \alpha}{\pi} \right)^{1/4}
 \int_{-\infty}^{+\infty} dE e^{-\alpha(E-E_0)^2- i s |E|} \Psi_{E}(Q) \nonumber \\
&\simeq&  \left(  \frac{2 \alpha}{\pi} \right)^{1/4}   \sqrt{ \frac{\lambda\sqrt{G}}{4\pi \sin(\lambda Q)} }
\int_{-\infty}^{+\infty} dE e^{-\alpha(E-E_0)^2- i s E}e^{i E x},
\label{PsiApprox}
\end{eqnarray}
where $x := \frac{1}{2} \sqrt{G}  \ln \left| \tan\left( \frac{\lambda Q}{2} \right)\right|$. The approximation
is valid because, for $E_0\sqrt{\alpha} \gg 1$, contribution from the negative energies to
integral (\ref{PsiApprox}) is negligible. Therefore, $|E|$ can be replaced by $E$. Calculating
the integral (\ref{PsiApprox}) we get
\begin{equation}\label{sP}
\Psi(Q,s) =  \sqrt{\frac{\lambda}{\sqrt{8\pi \tilde{\alpha}}}}
\frac{1}{\sqrt{\sin{(\lambda Q)}}}
e^{-\frac{(x-s)^2}{4\alpha}}e^{i E_0(x-s)},
\end{equation}
where $\tilde{\alpha} := \alpha/G$. The probability density is
easily found to be
\begin{equation}
\frac{|\Psi(Q,s)|^2}{\lambda} =
\frac{ \exp\left\{-\frac{1}{2\tilde{\alpha}}
\left[ \frac{1}{2} \ln \left| \tan\left( \frac{\lambda Q}{2} \right)
\right|-\frac{s}{t_{\rm{Pl}}}\right]^2\right\}
}{\sqrt{8\pi \tilde{\alpha}} \sin{(\lambda Q)}}.
\end{equation}

It is worth to emphasis  that the  state (\ref{sP}) is {\it
normalizable} so the probabilistic interpretation can be applied.
Thus, the function $|\Psi(Q,s)|^2$ gives the probability density
of finding universe with given $Q$ at the particular moment of
time $s$.  In the case $\lambda \rightarrow 0$, the state becomes
{\it non-normalizable}  and the probabilistic interpretation
cannot be applied.  Finding a normalizable state is a serious
problem  in  most of quantum cosmologies different from LQC
(see, e.g. \cite{CK}).

\subsection{Evolution of observable $\hat{Q}$}

\begin{figure}[ht!]
\centering
\includegraphics[width=10cm,angle=0]{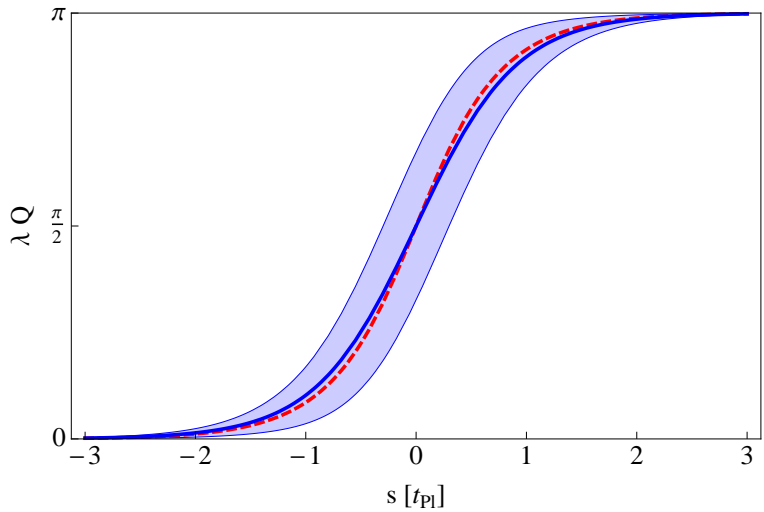}
\caption{The thick blue line  represents the mean value ${\langle
\hat{Q} \rangle}$ for $\tilde{\alpha}=0.1$. The dashed red line is
the classical solution $Q = \frac{2}{\lambda} \arctan \exp \left\{
2s/t_{\rm{Pl}} \right\}$. The shadowed region is constrained by
${\langle \hat{Q} \rangle} \pm \Delta\hat{Q}$.} \label{Fig1}
\end{figure}

One can now investigate the mean value of the $\hat{Q}$ operator
in the $|\Psi(s)\rangle$ state.  It is clear that  $\hat{Q}$ is a
symmetric bounded operator on $|\Psi(s)\rangle$. We find
\begin{eqnarray}
\label{expectQ}
{\langle \hat{Q} \rangle} := \langle \Psi(s) |\hat{Q}| \Psi(s)
\rangle \nonumber \\
= \frac{1}{\sqrt{2\pi \tilde{\alpha}}} \frac{2}{\lambda}
\int_{-\infty}^{+\infty}   \arctan\left\{ \exp \left[ 2\left( y +
\frac{s}{t_{\rm{Pl}}}\right)  \right] \right\} \exp \left\{-
\frac{1}{2\tilde{\alpha}}y^2 \right\} dy.
\end{eqnarray}
One can show that
\begin{equation}
\lim_{s\rightarrow +\infty} \lambda \langle \hat{Q} \rangle (s) =
\pi,~~~~~~ \lim_{s\rightarrow -\infty} \lambda \langle \hat{Q}
\rangle (s) = 0,
\end{equation}
which agrees with the classical limits. We also find
\begin{eqnarray}
\label{expectQ2}
\langle \Psi(s) |\hat{Q}^2| \Psi(s) \rangle   = \nonumber \\
= \frac{1}{\sqrt{2\pi
\tilde{\alpha}}} \left(\frac{2}{\lambda}\right)^2
\int_{-\infty}^{+\infty}   \arctan^2\left\{ \exp \left[ 2\left( y
+ \frac{s}{t_{\rm{Pl}}} \right)  \right] \right\}  e^{-
\frac{1}{2\tilde{\alpha}}y^2 } dy.
\end{eqnarray}
Thus, the {\it dispersion} of $\hat{Q}$ in the state
$|\Psi(s)\rangle$ reads\footnote{The integrals (\ref{expectQ}) and
(\ref{expectQ2}) have been determined numerically.}
\begin{equation}
\Delta \hat{Q} := \sqrt{ \langle \Psi(s) |\hat{Q}^2| \Psi(s) \rangle-(\langle
\Psi(s) |\hat{Q}| \Psi(s) \rangle)^2}.
\end{equation}
In Fig. \ref{Fig1} we show ${\langle \hat{Q} \rangle}$ (thick blue line) and compare it
with the classical solution (dashed red line)
\begin{equation}
Q = \frac{2}{\lambda} \arctan \exp \Big(
\frac{2}{\sqrt{G}}(T-T_0)\Big),
\end{equation}
with $s=T-T_0$.
The shadowed region represents the dispersion of our state, and it
is constrained by ${\langle \hat{Q} \rangle} \pm \Delta\hat{Q}$.
One can notice that, close to the bounce at $s=0$ (where $\lambda
Q =\pi/2$), the dispersion of the state is significant, while it
fast decreases as $s \rightarrow \pm \infty$. Therefore, far from
the bounce, the state converge to the classical
solution\footnote{ One can  notice some little discrepancy between
the classical solution and the  mean value ${\langle \hat{Q}
\rangle}$. This is however due to the particular form of the state
(\ref{gauss}) that has been chosen.}.

\subsection{Evolution of observable $\hat{P}$}

The volume operator, due to (\ref{redpar}), reads
\begin{eqnarray}
\hat{v} = 4\pi l_{\rm{Pl}}^2 \gamma \hat{P}.
\end{eqnarray}
The $\hat{P}$ operator is unbounded, but  symmetric on the state
$|\Psi(s)\rangle$ defined by Eq. (\ref{sP}):
\begin{eqnarray}
\langle \Psi(s) |\hat{P} \Psi(s) \rangle = -i \frac{\lambda}{\sqrt{8\pi
\tilde{\alpha}}}
\lim_{x\rightarrow +\infty} \cosh\left(\frac{2}{\sqrt{G}}x\right)
\left[ e^{-\frac{(x-s)^2}{2\alpha}}-e^{-\frac{(x+s)^2}{2\alpha}}  \right]\nonumber \\
+\langle \hat{P}\Psi(s)|\Psi(s) \rangle
=\langle \hat{P}\Psi(s)|\Psi(s) \rangle,
\label{symmP}
\end{eqnarray}
where we used the relation $\cosh\left(\frac{2}{\sqrt{G}}x\right)
= \frac{1}{\sin(\lambda Q)}$.

\begin{figure}[ht!]
\centering
\includegraphics[width=10cm,angle=0]{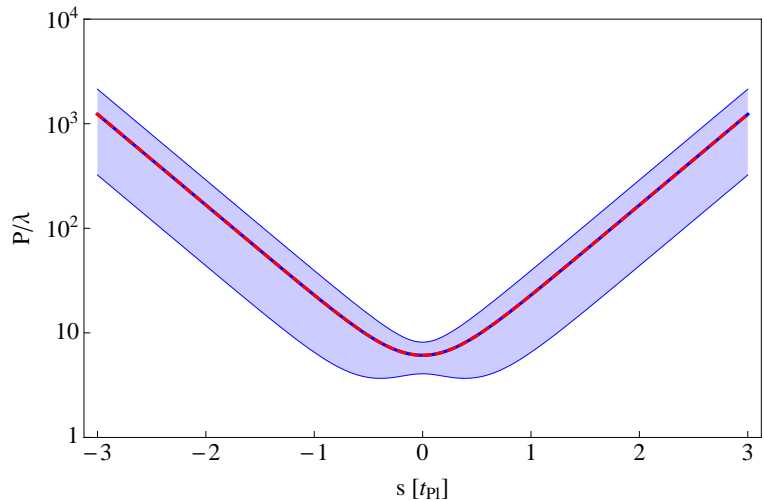}
\caption{The thick blue line  represents the mean value ${\langle
\hat{P} \rangle}$ for $\tilde{\alpha}=0.1$ and $\tilde{E}_0=10$.
The shadowed region is constrained by
${\langle \hat{P} \rangle} \pm \Delta\hat{P}$.}
\label{Fig2}
\end{figure}

It is not difficult to  derive the following:
\begin{eqnarray}
{\langle \hat{P} \rangle} := \langle \Psi(s) |\hat{P}| \Psi(s)\rangle  \nonumber \\
= \frac{\lambda E_0 }{\sqrt{8 \pi \tilde{\alpha}}}
\int_{-\infty}^{+\infty}   \cosh \left( \frac{2}{\sqrt{G}}x \right)  \exp \left\{-
\frac{1}{2\alpha}(x-s)^2 \right\} dx   \nonumber \\
 = \frac{1}{2} \lambda E_0 l_{\rm{Pl}}  e^{2\tilde{\alpha}} \cosh
 \left( \frac{2s}{\sqrt{G}} \right). \label{evoP}
\end{eqnarray}
Therefore,
\begin{eqnarray}
{\langle \hat{v} \rangle}= 4\pi l_{\rm{Pl}}^2 \gamma {\langle \hat{P} \rangle}
= 2\pi l_{\rm{Pl}}^3 \gamma \lambda E_0
e^{2\tilde{\alpha}} \cosh \left( \frac{2s}{\sqrt{G}} \right).
\end{eqnarray}
The minimum allowed volume in the $\Psi$ state is
\begin{eqnarray}
{\langle \hat{v} \rangle}_{\rm{min}} = 2\pi l_{\rm{Pl}}^3 \gamma \lambda E_0
e^{2\tilde{\alpha}}.
\end{eqnarray}
The corresponding classical solution reads
\begin{equation}
v = v_0 \cosh \Big(
\frac{2}{\sqrt{G}}(T-T_0)\Big).
\end{equation}
Thus, the functional forms  of $\langle \hat{v} \rangle$ and
$v$ coincide. We also find
\begin{eqnarray}
\langle \Psi(s) |\hat{P}^2| \Psi(s) \rangle = \frac{\lambda^2}{32}
\left[ \left(\frac{1}{\tilde{\alpha}}-4+4\tilde{E}_0^2 \right)  \right. \nonumber \\
+ \left. \cosh\left(\frac{4s}{\sqrt{G}} \right)e^{8\tilde{\alpha}}\left(\frac{1}{\tilde{\alpha}}+4+4\tilde{E}_0^2\right)    \right],
\end{eqnarray}
where $\tilde{E}_0 := E_0/E_{\rm{Pl}}$. The dispersion of
$\hat{P}$ in the state $|\Psi(s)\rangle$ is found to be
\begin{eqnarray}
\Delta \hat{P} &:=& \sqrt{ \langle \Psi(s) |\hat{P}^2| \Psi(s) \rangle-(\langle
\Psi(s) |\hat{P}| \Psi(s) \rangle)^2}  \nonumber \\
         &=& \frac{\lambda}{8} \left[ 2\left(\frac{1}{\tilde{\alpha}}-4+4\tilde{E}_0^2
         \right)+2\cosh\left(\frac{4s}
{\sqrt{G}} \right)e^{8\tilde{\alpha}}\left(\frac{1}{\tilde{\alpha}}+4+4\tilde{E}_0^2
\right)  \right. \nonumber \\
&-& \left. 16 \tilde{E}^2_0 e^{4\tilde{\alpha}}\cosh^2\left(\frac{2s}{\sqrt{G}} \right) \right]^{1/2}
\end{eqnarray}
We visualize the dispersion of $\hat{P}$ as a function of time in Fig. \ref{Fig2}.\\

\noindent We can see that both the volume and its dispersion grow
quite fast away (exponentially) from the big bounce
region\footnote{The $P$-axis uses the logarithmic scale. }. This
can be understood in the context of Heisenberg's relation. Namely,
while dispersion $\Delta \hat{Q}$ tends to zero, for $s
\rightarrow \pm \infty$, the dispersion $\Delta \hat{P}$ grows
appropriately to fulfill uncertainty relation $\Delta
\hat{Q}\Delta \hat{P} \geq 1/2$. We shall study this issue in more
details in the next subsection.

\subsection{The Heisenberg uncertainty relation}

The  standard probabilistic interpretation of quantum mechanics
cannot be applied to a cosmological system for a number of
reasons. For instance, there is only {\it one} Universe and there
is {\it no} observer outside the Universe.  Thus, the process of
measurement in quantum cosmology may differ from that known from
the Copenhagen interpretation of quantum mechanics. Instead of
complaining at the interpretation problems, it makes sense
verification if some fundamental relations {\it underlying}
quantum mechanics are satisfied. The best known is the Heisenberg
uncertainty relation. What is its status in our cosmological
setup? Is it satisfied during the quantum big bounce transition?

Our two canonical variables $\hat{Q}$ and $\hat{P}$ do not
commute: $[\hat{Q},\hat{P}] = i \hat{\mathbb{I}}$. Since both
operators are {\it symmetric} on the space of states
$|\Psi(s)\rangle$, they should satisfy algebraically the Heisenberg
relation:
\begin{equation}\label{relH}
\Delta\hat{Q}\;\Delta\hat{P} \geq 1/2.
\end{equation}

\begin{figure}[ht!]
\centering
\includegraphics[width=10cm,angle=0]{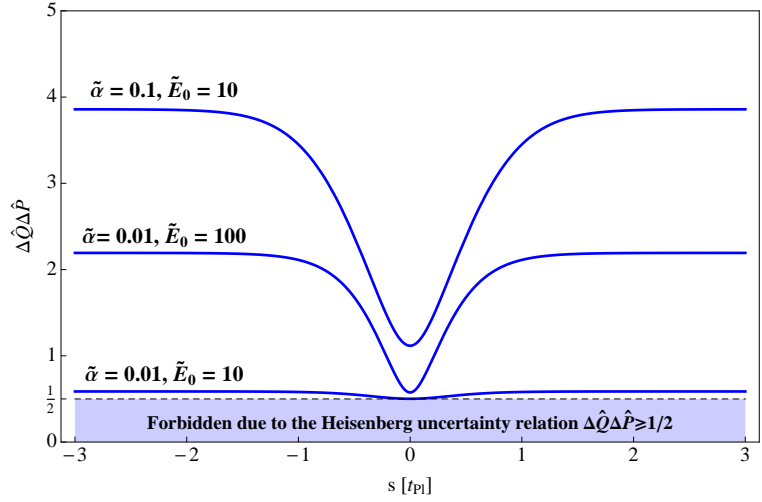}
\caption{Evolution of the product of dispersions $\Delta
\hat{Q}\Delta \hat{P}$. Heisenberg's uncertainty relation $\Delta
\hat{Q}\Delta \hat{P} \geq 1/2 $ is preserved during the
evolution. } \label{Fig3}
\end{figure}

In Fig. \ref{Fig3}  we show the evolution of $\Delta \hat{Q}\Delta
\hat{P}$ for the model considered in this paper.
We can see that the Heisenberg relation (\ref{relH}) is perfectly
satisfied during the entire evolution, for any values of the
parameters  $\alpha$ and $E_0$ of the state. However, one
cannot ascribe {\it probabilistic} interpretation to this
relation. It is so because the representation $[\hat{Q},\hat{P}] =
i \hat{\mathbb{I}}$ cannot be self adjoint owing to the fact
(shown earlier) that $\hat{Q}$ and $\hat{P}$ are bounded and
unbounded operators, respectively \cite{CRP}.

The function $\Delta \hat{Q}\Delta \hat{P}$ is  symmetric with
respect to the bounce and reaches its minimal value at the bounce.
This is rather surprising result. Namely, one would expect that
the transition point ($s=0$) is the most \emph{quantum} part of
the evolution, while in the limits $s \rightarrow \pm \infty$ one
should get the most \emph{classical} evolution.  The situation is
however quite the opposite. The transition point (big bounce) is
the  least quantum  part of the evolution! In turn, in the limits
$s \rightarrow \pm \infty$ the product of dispersions $\Delta
\hat{Q}\Delta \hat{P}$ is saturated:
\begin{eqnarray}
\lim_{s\rightarrow \pm \infty} \Delta\hat{Q}\Delta \hat{P} = \nonumber \\
 = \frac{1}{2} e^{4\tilde{\alpha}} \tilde{E}_0 \sqrt{
e^{4\tilde{\alpha}}-1} \sqrt{
e^{4\tilde{\alpha}}-1+e^{4\tilde{\alpha}} \left(
\frac{1}{4\tilde{E}_0^2\tilde{\alpha} }+
\frac{1}{\tilde{E}_0^2}\right) } \geq \frac{1}{2},
\label{Hr}
\end{eqnarray}
which shows that the  Gaussian packet we consider is  always
quantum.  We can also see that the relation (\ref{Hr}) {\it does
not} depend on the parameter $\lambda$, which is a remarkable
feature of our quantization scheme\footnote{The parameter
$\lambda$ appears in the formalism as the result of approximating
the curvature of connection by holonomies around small loops with
length $\lambda$. It is a free parameter of the nonstandard LQC
and fixed parameter of the standard LQC, so it is of basic
importance.}

\begin{figure}[ht!]
\centering
\includegraphics[width=10cm,angle=0]{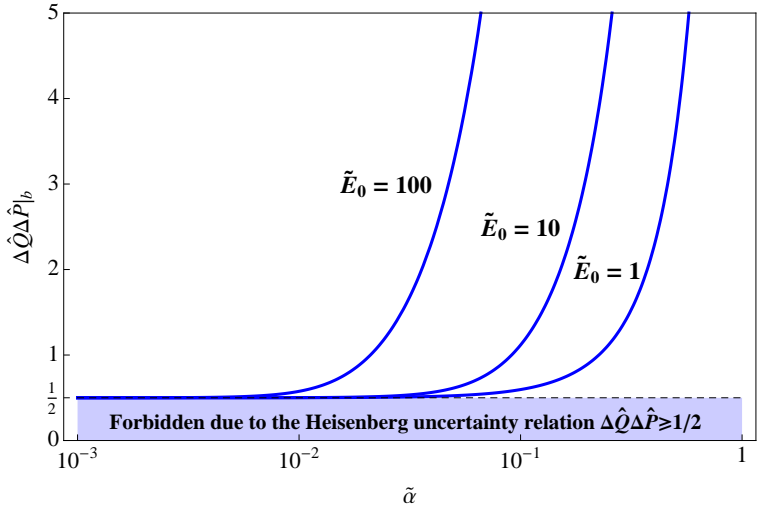}
\caption{The value of $\Delta\hat{Q}\Delta \hat{P}$ at the bounce
($s=0$) as a function of the parameters of state.}
\label{Fig4}
\end{figure}

\begin{figure}[ht!]
\centering
\begin{tabular}{cc}
{\bf a)}  \includegraphics[width=7cm,angle=0]{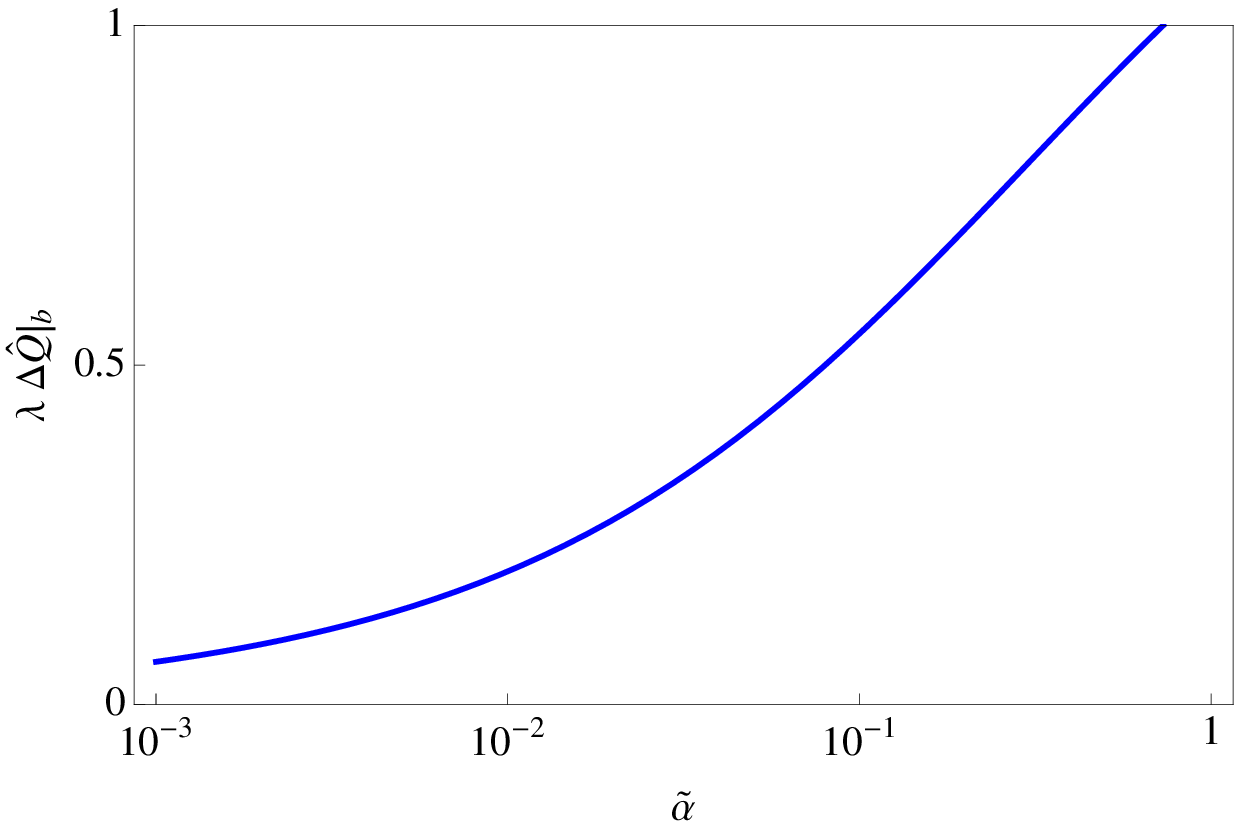} &
{\bf b)} \includegraphics[width=6.8cm,angle=0]{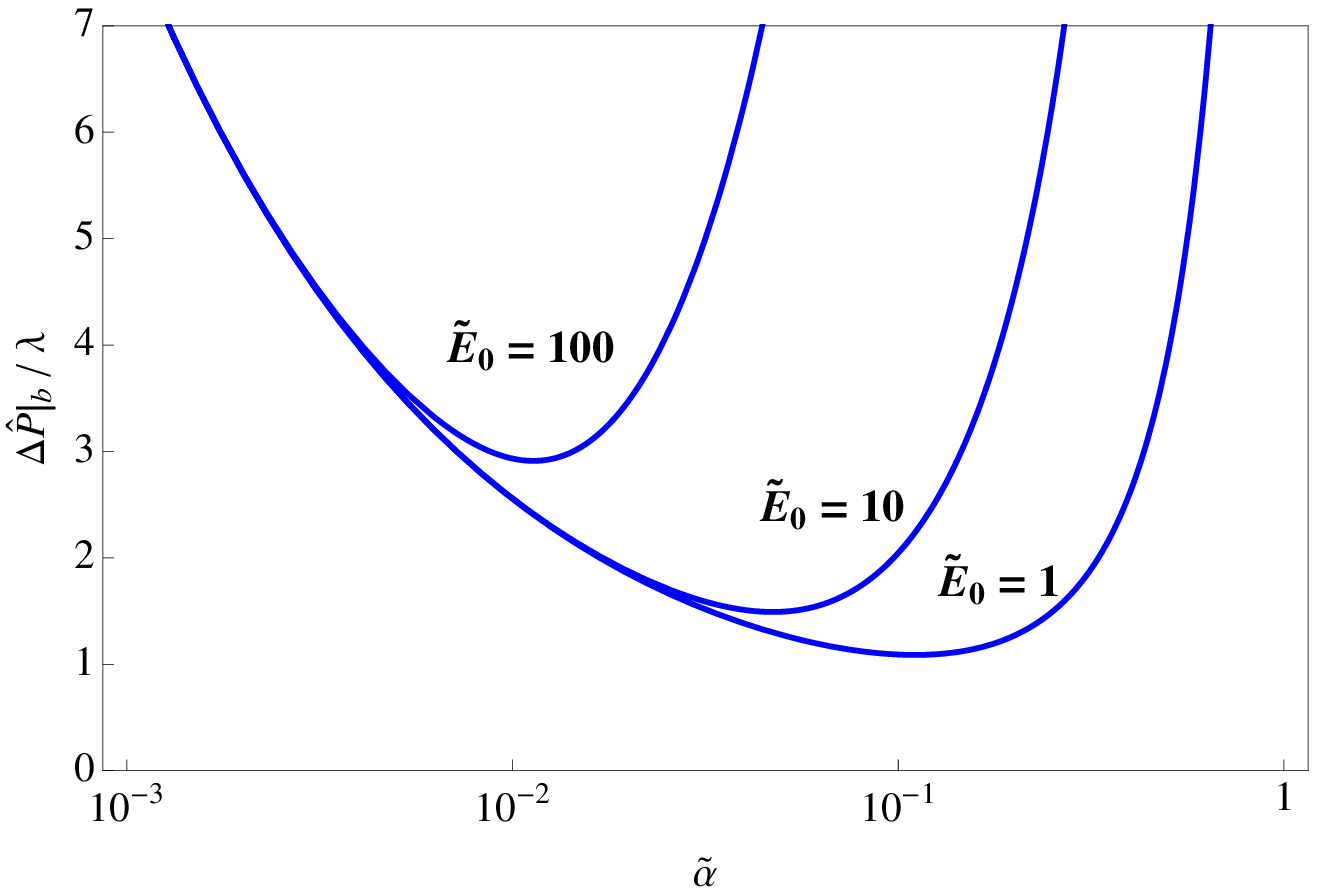}
\end{tabular}
\caption{{\bf a)} Dispersion $\lambda \Delta \hat{Q}$ at the bounce
$(s=0)$ as a function of the parameter $\tilde{\alpha}$. {\bf b)} Dispersion
$\Delta \hat{P} / \lambda$ at the bounce
$(s=0)$ as a function of the parameter $\tilde{\alpha}$ for $\tilde{E}_0=1,10$ and $100$. }
\label{Fig5}
\end{figure}

Let us investigate in more details the value of $\Delta
\hat{Q}\Delta \hat{P}$ at the bounce, $\Delta \hat{Q}\Delta
\hat{P}|_{\rm{b}}$. In Fig. \ref{Fig4} we show $\Delta
\hat{Q}\Delta \hat{P}|_{\rm{b}}$ as a function of
$\tilde{\alpha}$ for $\tilde{E}_0=1,10$ and $100$.
 As we can see, the boundary value $\Delta \hat{Q}\Delta
\hat{P}|_{\rm{b}}=1/2$ is never crossed and it is approached for
$\tilde{\alpha} \rightarrow 0$. Therefore, the more sharply peaked
the state is, the smaller value of $\Delta \hat{Q}\Delta \hat{P}$
is at the bounce.  For any value of $\tilde{\alpha}$, the smaller
value of $\tilde{E}_0$ is  the boundary $\Delta \hat{Q}\Delta
\hat{P}|_{\rm{b}}=1/2$ is easier approached. In order to
understand this dependence better, let us investigate separately
$\Delta \hat{Q}|_{\rm{b}} $ and $\Delta \hat{P}|_{\rm{b}}$. We
show these dispersions, as a function of $\tilde{\alpha}$, in the
left and right panels of Fig. \ref{Fig5}, respectively.
The function $\Delta \hat{Q}|_{\rm{b}}$ grows monotonically with
increase of $\tilde{\alpha}$ and is independent on $\tilde{E}_0$.
The dependence on $\tilde{\alpha}$ is more complex for $\Delta
\hat{P}|_{\rm{b}}$. Namely for any $\tilde{E}_0$ there always
exists some $\tilde{\alpha}$ at which the function $\Delta
\hat{P}|_{\rm{b}}(\tilde{\alpha})$ takes the minimum. The
smallest possible value of $\Delta \hat{P}|_{\rm{b}} \approx
1.0686 \lambda$ is reached for $\tilde{E}_0 \rightarrow 0$ and
$\tilde{\alpha} \approx 0.11797$.

In summary, our quantum cosmological setup is devoid of {\it
complete} standard  probabilistic interpretation, but satisfies
formally the most basic relationships of the standard quantum
mechanics.

\subsection{Problem of time}

An evolution of classical variables $Q$  and $P$, as presented by
Eqs. (\ref{vQ}) and (\ref{vP}), is parametrized by a free massless
scalar field $\phi$, due to (\ref{redpar}), that is a monotonic
function  so it can play the role of an internal clock
\cite{Dzierzak:2009ip}. The expectation values of the
corresponding quantum operators $\hat{Q}$ and $\hat{P}$, defined
by (\ref{evo}) and (\ref{evoP}), are parametrized by $s \in \dR$,
owing to (\ref{evo}). Are the evolution variables $\phi$ and $s$
quite independent? An important difference between them is that
they label an evolution of the system at different levels:
classical and quantum, respectively. We {\it postulate} that these
two variables are {\it related} linearly:  $s = a_1 \phi + a_2,$
where $a_1,a_2 \in \dR.$ It means that neither $\phi$ nor $s$
belong to the physical phase space. This seems to be the specific
feature of our reduced phase space (RPS) method. It has been
already proposed in \cite{Malkiewicz:2009zd} treating the $\phi$
variable as an evolution parameter  of {\it both} classical and
quantum dynamics. Such an interpretation is supported by the plots
of Fig. \ref{Fig1} and Fig. \ref{Fig2}, where the same abscissa is
used to label both classical and quantum evolution of presented
functions. The plots of classical and corresponding quantum
functions practically {\it coincide}. Such an agreement suggests
that our postulate concerning time variable is reasonable.

\section{Relative fluctuations}

 In this section we  study the relative fluctuations of the
quantum observables $\hat{\mathcal{O}}$  in the state $|\Psi
\rangle$. We  consider three observables: $\hat{\mathbb{H}}$,
$\hat{Q}$ and $\hat{P}$. The relative fluctuation $\Delta
\hat{\mathcal{O}}/{\langle \hat{\mathcal{O}} \rangle}$ is a
measure of the  semi-classicality of a quantum state. We say that
$|\Psi \rangle$ is  {\it semiclassical} if $\;\Delta
\hat{\mathcal{O}} /{\langle \hat{\mathcal{O}}\rangle} \ll 1$, and
{\it quantum} if $\;\Delta\hat{\mathcal{O}}/\langle
\hat{\mathcal{O}} \rangle \sim 1$.  It is clear that the
simiclassicality notion is not at all defined uniquely. We apply
such a definition of semiclassicality because we wish to be able
to make comparison of our analyzes with the available published
results \cite{BojowaldNature,Corichi:2007am,Bojowald:2007gc,
Corichi:2011rt,Kaminski:2010yz}. In the future we shall try to
apply a more sophisticated definition: If the uncertainty in an
observable is less than the observational {\it  precision}, the
state is semiclassical with respect to that observable.

In what follows, we  also consider the function
$D_{\mathcal{O}}$ characterizing the asymptotic aspects of
relative fluctuations with respect to the bounce, defined to be
\cite{Corichi:2007am}
\begin{equation}\label{diff}
D_{\mathcal{O}} := \lim_{s\rightarrow \infty} \left[ \left(
\frac{\Delta \hat{\mathcal{O}}(-s) }{{\langle \hat{\mathcal{O}}
\rangle}(-s)} \right)^2 -  \left( \frac{\Delta
\hat{\mathcal{O}}(s) }{{\langle \hat{\mathcal{O}} \rangle}(s)}
\right)^2 \right].
\end{equation}

\subsection{Relative fluctuations of $\hat{\mathbb{H}}$}

Let us start from deriving expectation value of $\hat{\mathbb{H}}$
in the state $ |\Psi(s)\rangle$, we find
\begin{equation}
{\langle \hat{\mathbb{H}} \rangle} := \langle \Psi(s)
|\hat{\mathbb{H}}| \Psi(s) \rangle = E_0 \;\rm{erf} \left(
\sqrt{2 \alpha} E_0\right) +\frac{e^{-2 \alpha E_0^2}}{\sqrt{2\pi
\alpha}},
\end{equation}
where $\rm{erf}(x):= \frac{2}{\sqrt{\pi}} \int_0^xe^{-t^2}dt$ is the error function. One can
see that for $\sqrt{\alpha} E_0 \gg 1$, the above expression
simplifies to ${\langle \hat{\mathbb{H}} \rangle} \simeq E_0$. In
order to find the dispersion of $\hat{\mathbb{H}}$, we also need:
\begin{equation}
 \langle \Psi(s) |\hat{\mathbb{H}}^2| \Psi(s)
\rangle =  \frac{1}{4\alpha} +E_0^2.
\end{equation}
Based on the above, we determine
\begin{eqnarray}
\Delta \hat{\mathbb{H}} := \sqrt{ \langle \Psi(s) |\hat{\mathbb{H}}^2| \Psi(s)
\rangle-(\langle
\Psi(s) |\hat{\mathbb{H}}| \Psi(s) \rangle)^2}  \nonumber \\
         =  \left\{ \frac{1}{4\alpha}+E_0^2\left[1-\rm{erf}^2 \left(
         \sqrt{2 \alpha} E_0\right)\right] \right. \\
 \left.    -\frac{2 E_0 e^{-2 \alpha E_0^2}}{\sqrt{2\pi \alpha}} \rm{erf}
          \left(  \sqrt{2 \alpha} E_0\right)
         -\frac{e^{-4 \alpha E_0^2}}{2\pi \alpha} \right\}^{1/2}.
\end{eqnarray}
We can see that for  $ \sqrt{\alpha} E_0 \gg 1$, dispersion of
$\hat{\mathbb{H}}$ simplifies to $\Delta \hat{\mathbb{H}} \simeq
1/\sqrt{4\alpha}$. Therefore, for  $ \sqrt{\alpha} E_0 \gg 1$, the
relative dispersion
\begin{equation}
\frac{\Delta \hat{\mathbb{H}} }{\langle \hat{\mathbb{H}} \rangle}
\simeq \frac{1}{2 \sqrt{\alpha} E_0}  \ll 1.
\label{RelHsemclass}
\end{equation}
It is worth to note that condition $ \sqrt{\alpha} E_0 \gg 1$ was
also used while performing integration  (\ref{PsiApprox}). One can
see now that approximation based on this condition is justified by
the restriction imposed on the relative fluctuations of the
Hamiltonian $\hat{\mathbb{H}}$. The semiclassicality requires $
\sqrt{\alpha} E_0 \gg 1$. The relative fluctuations of
$\hat{\mathbb{H}}$ are {\it constant} in time and therefore
symmetric with respect to the bounce so finally we get:
$D_{\mathbb{H}} = 0$.

\subsection{Relative fluctuations of $\hat{P}$}

\begin{figure}[ht!]
\centering
\begin{tabular}{cc}
{\bf a)} \includegraphics[width=6.8cm,angle=0]{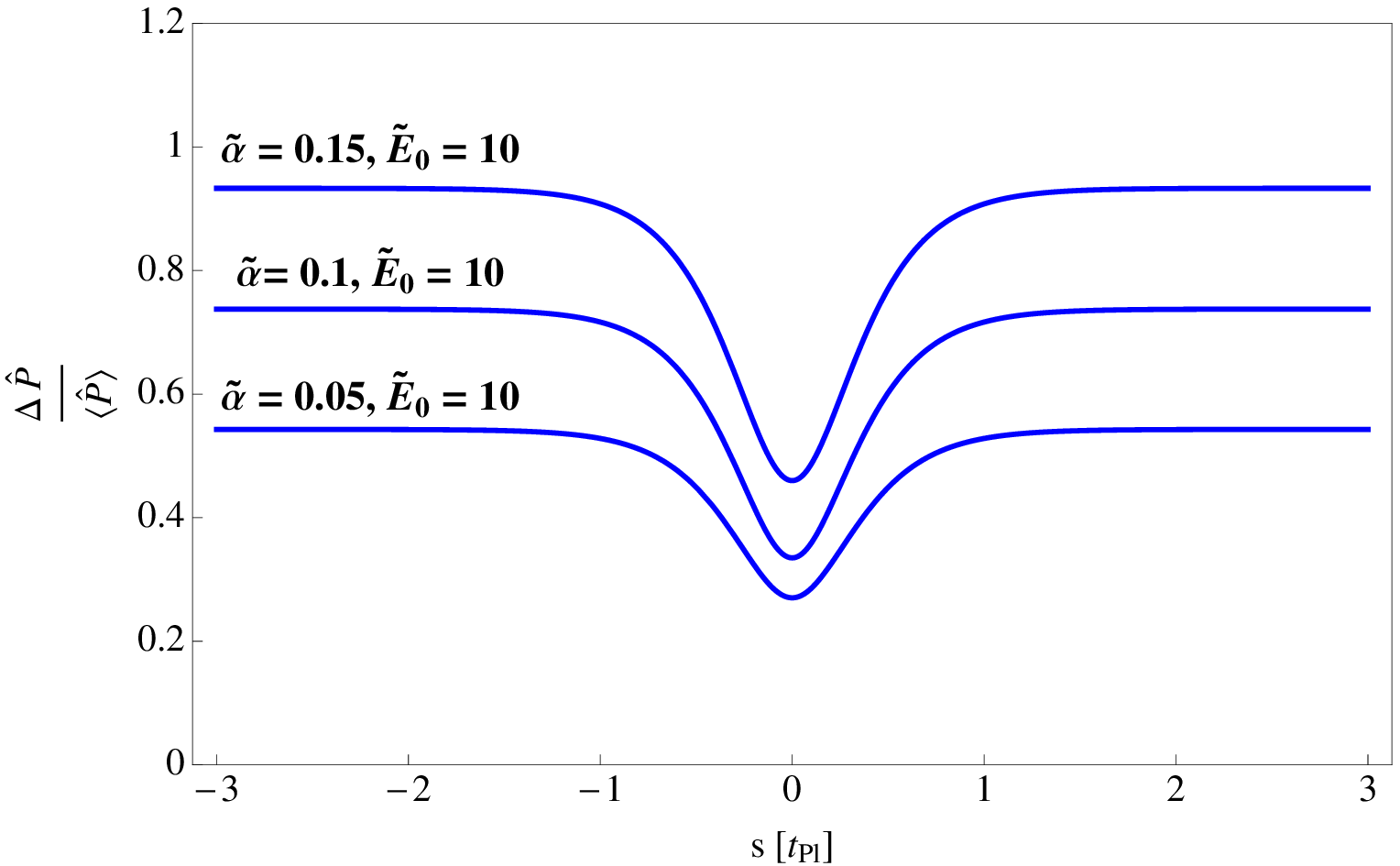} &
{\bf b)} \includegraphics[width=7.1cm,angle=0]{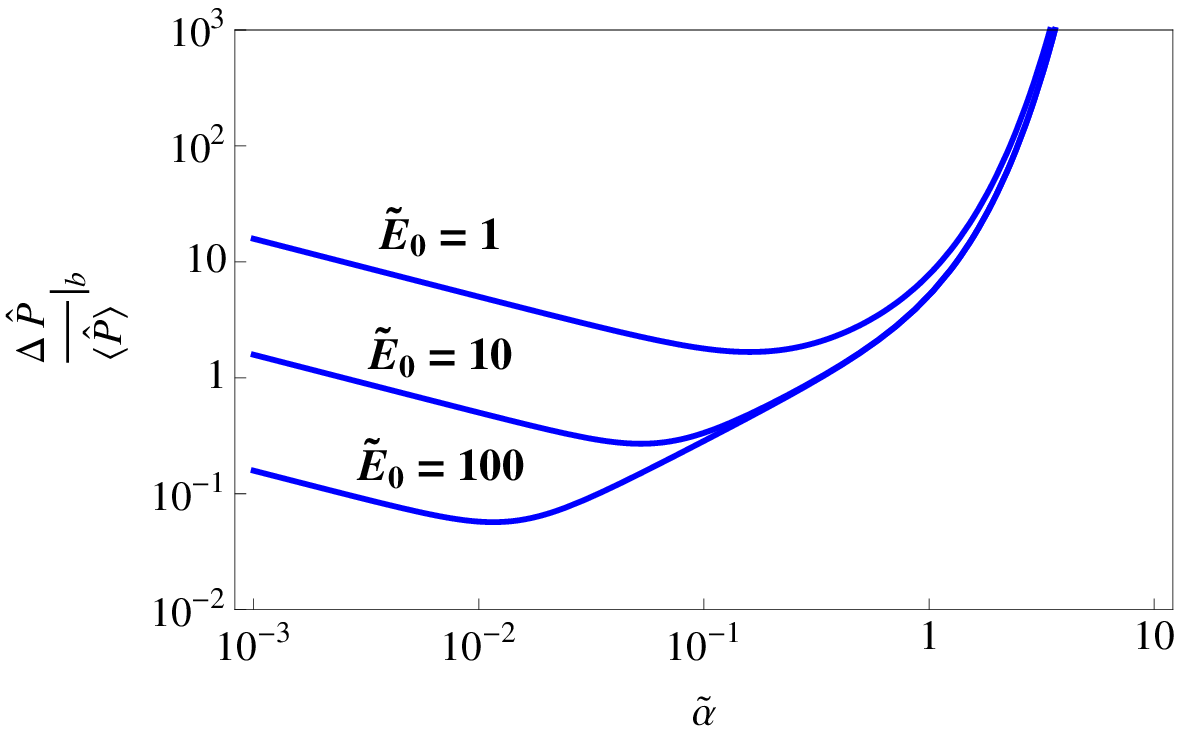}
\end{tabular}
\caption{ {\bf a)} Relative fluctuations  $\frac{\Delta \hat{P}}{{\langle
\hat{P} \rangle}}$ for different values of $\tilde{\alpha}$ and $\tilde{E}_0$.
{\bf b)}  Relative fluctuations $\frac{\Delta \hat{P}}{\langle \hat{P}
\rangle } $ at the bounce ($s=0$) as a function of  $\tilde{\alpha}$ for
$\tilde{E}_0= 1, 10$ and $100$. }
\label{Fig6}
\end{figure}

Relative fluctuations of  $\hat{P}$ can be expressed as follows
\begin{equation}\label{fP}
\frac{\Delta \hat{P}}{\langle \hat{P} \rangle} =  \sqrt{
\frac{\left(\frac{1}{\tilde{\alpha}}-4+4\tilde{E}_0^2 \right)
+\cosh\left(\frac{4s}{\sqrt{G}}
\right)e^{8\tilde{\alpha}}\left(\frac{1}{\tilde{\alpha}}
+4+4\tilde{E}_0^2 \right)}{8\tilde{E}_0^2e^{4\tilde{\alpha}}
\cosh^2 \left( \frac{2s}{\sqrt{G}} \right)}-1}.
\end{equation}
In the left part of Fig. \ref{Fig6} we plot this relation for the
different values of $\tilde{\alpha}$ and $\tilde{E}_0$.
The relative fluctuations of $\hat{P}$ are symmetric with respect
to the bounce and reach the minimal  value at the transition point
between the contracting and expanding phases.  The symmetry $s
\rightarrow -s $ directly implies that
\begin{equation}
D_P = \lim_{s\rightarrow \infty} \left[ \left(  \frac{\Delta
\hat{P}(-s) }{{\langle \hat{P} \rangle}(-s)} \right)^2 -  \left(
\frac{\Delta \hat{P}(s) }{{\langle \hat{P} \rangle}(s)} \right)^2
\right] = 0 .
\end{equation}
Thus, at infinite past and future times, the relative fluctuations
of $\hat{P}$ are the same:
\begin{equation}
\left.\frac{\Delta \hat{P} }{\langle\hat{P} \rangle}
\right|_{\rm{max}} =\lim_{s\rightarrow \pm \infty} \frac{\Delta
\hat{P}}{\langle \hat{P}\rangle} =
\sqrt{e^{4\tilde{\alpha}}-1+e^{4\tilde{\alpha}}\left(
\frac{1}{4\tilde{E}_0^2\tilde{\alpha} }
+\frac{1}{\tilde{E}_0^2}\right)}.
\end{equation}
The function  $\left.\frac{\Delta \hat{P} }{\langle\hat{P}
\rangle} \right|_{\rm{max}} (\tilde{\alpha})$ has the minimum
for any  $\tilde{E}_0$, which  is located at
\begin{equation}
\tilde{\alpha}_{\rm{min}} := \frac{-1+\sqrt{5+4\tilde{E}_0^2
}}{8(1+\tilde{E}_0^2)}. \label{alfamin}
\end{equation}
Therefore, having $\tilde{E}_0$, one can always minimize $\left.
\frac{\Delta \hat{P} }{\langle\hat{P} \rangle}
\right|_{\rm{max}} (\tilde{\alpha})$ by choosing
$\tilde{\alpha}= \tilde{\alpha}_{\rm{min}}$. If $\tilde{E}_0 \gg
1$, the expression (\ref{alfamin}) can be approximated by
$\tilde{\alpha}_{\rm{min}} \approx 1/(4\tilde{E}_0).$

Let us now consider the fluctuations of $\hat{P}$ at the bounce,
which can be expressed as follows
\begin{equation}
\left.  \frac{\Delta \hat{P}}{\langle \hat{P} \rangle}
\right|_{\rm{b}} =  \sqrt{
\frac{\left(\frac{1}{\tilde{\alpha}}-4+4\tilde{E}_0^2 \right)
+e^{8\tilde{\alpha}}\left(\frac{1}{\tilde{\alpha}}
+4+4\tilde{E}_0^2 \right)}{8\tilde{E}_0^2e^{4\tilde{\alpha}}}-1} .
\end{equation}
We plot this function in the right part of  Fig. \ref{Fig6}  for
fixed values of $\tilde{E}_0=1, 10$ and $100$. As we can see, for
any value of $\tilde{E}_0$, there is some $\tilde{\alpha}$ at
which fluctuations take the minimum. These minimal value
fluctuations decrease with the increase of $\tilde{E}_0$. In the
limit $\tilde{E}_0 \rightarrow \infty $, the relative fluctuations
at the bounce  are given by
\begin{equation}
\lim_{\tilde{E}_0 \rightarrow \infty} \left.\frac{\Delta \hat{P}
}{\langle\hat{P} \rangle} \right|_{\rm{b}} = \sqrt{2}
\sinh(2\tilde{\alpha}).
\end{equation}
Therefore, the fluctuations $\hat{P}$  at the bounce go to zero
for $\tilde{\alpha} \rightarrow 0$ and   $\tilde{E}_0 \rightarrow
\infty $.

\subsection{Relative fluctuations of $\hat{Q}$}

We have found (see, Appendix A) that after the bounce the relative
fluctuations are decreasing and go to zero, so we have
\begin{equation}
\lim_{s\rightarrow +\infty} \frac{\Delta \hat{Q}(s) }{{\langle
\hat{Q} \rangle}(s)}=  0. \label{relQpinf}
\end{equation}
Therefore, the state becomes a semiclassical one. However, it may
not be the case in the far past for  large enough
$\tilde{\alpha}$. While moving backward in time the relative
fluctuations saturate. This saturated value can be found (see,
Appendix B) by calculating the integrals: $\langle \Psi(s)
|\hat{Q}| \Psi(s) \rangle$ and $\langle \Psi(s) |\hat{Q}^2|
\Psi(s) \rangle$ . One gets
\begin{eqnarray}\label{relQ}
\left.\frac{\Delta \hat{Q} }{\langle\hat{Q} \rangle} \right|_{\rm{max}} =
\lim_{s\rightarrow -\infty}  \frac{\Delta \hat{Q}(s) }{{\langle
\hat{Q} \rangle}(s)} = \sqrt{e^{4\tilde{\alpha}}-1},
\label{relQminf}
\end{eqnarray}
where $\tilde{\alpha} > 0$. Thus, before the bounce the state may
be a quantum one if the value $\tilde{\alpha}$ is sufficiently
large.

\begin{figure}[ht!]
\centering
\includegraphics[width=10cm,angle=0]{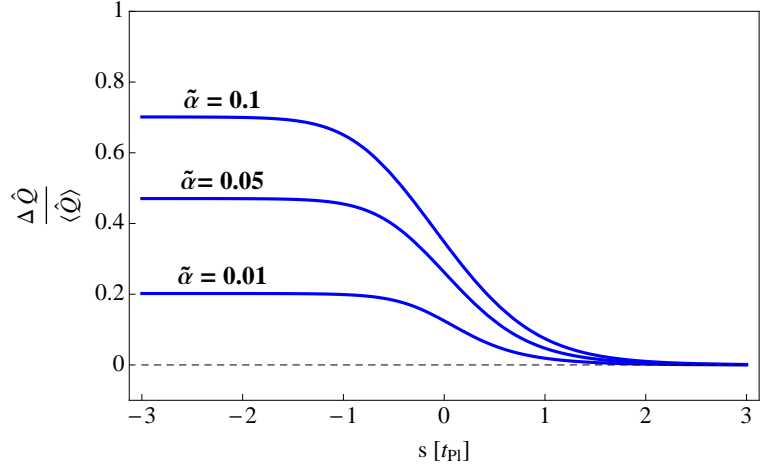}
\caption{  Relative fluctuations  $\frac{\Delta
\hat{Q}}{{\langle \hat{Q} \rangle}}$ for different values of
$\tilde{\alpha}$}
\label{Fig7}
\end{figure}

In Fig. \ref{Fig7} we present the plot of $\Delta
\hat{Q}/{\langle \hat{Q} \rangle}$ for different values of
$\tilde{\alpha}$.
The relative fluctuations are not symmetric across the bounce
($s=0$). For $s\rightarrow +\infty$ the relative fluctuations
converge to zero. For $s\rightarrow -\infty$ they saturate giving
$\sqrt{e^{4\tilde{\alpha}}-1}$. The difference in the asymptotic
values of the relative fluctuations is found to be
\begin{equation}
D_Q = e^{4\tilde{\alpha}}-1.
\end{equation}
In order o interpret these results it is crucial to recall that
directions of the parameter of time $s$ and coordinate time $t$
are opposite. Therefore, the positive values of $s$ correspond to
contraction while negative to expansion. Therefore, the relative
fluctuations of $\hat{Q}$ grow from the contracting to the
expanding phase. So, the universe becomes more quantum with the
increase of time $t$. The fluctuations tends to zero for $
t\rightarrow -\infty$.  Therefore the universe started its
evolution from a sharply peaked state. Let us call it a
semiclassical state. Only if the value of $\tilde{\alpha}$ is
sufficiently small the universe will keep its semiclassicality
during the whole evolution.

\subsection{Semiclassicality}

One can find that the maximal relative fluctuations of $\hat{P}$
and $\hat{Q}$ are related  as follows
\begin{equation}
\left(\left. \frac{\Delta \hat{P}}{\langle \hat{P}\rangle} \right|_{\rm{max}} \right)^2=
\left(\left. \frac{\Delta \hat{Q}}{\langle \hat{Q}\rangle} \right|_{\rm{max}} \right)^2
+e^{4\tilde{\alpha}} \left( \frac{1}{4\tilde{E}_0^2\tilde{\alpha}
}+ \frac{1}{\tilde{E}_0^2}\right).
\end{equation}
Thus, we have
\begin{equation}
\left. \frac{\Delta \hat{P}}{\langle \hat{P}\rangle} \right|_{\rm{max}} \geq  \left.
\frac{\Delta \hat{Q}}{\langle \hat{Q}\rangle} \right|_{\rm{max}}.
\label{PmaxQmax}
\end{equation}
The equality is  obtained in the limit ${\tilde{E}_0 \rightarrow
\infty}$, which leads to the equation:
\begin{equation}
\left. \frac{\Delta \hat{P}}{\langle \hat{P}\rangle} \right|_{\rm{max}} =
\sqrt{e^{4\tilde{\alpha}} -1} =  \left. \frac{\Delta \hat{Q}}{\langle \hat{Q}\rangle}
\right|_{\rm{max}}.
\end{equation}
It is clear that the semiclassicality condition
\begin{equation}
\left. \frac{\Delta \hat{P}}{\langle \hat{P}\rangle} \right|_{\rm{max}} \ll 1,
\label{semiP}
\end{equation}
implicates, due to (\ref{PmaxQmax}), that we have
\begin{equation}
\left. \frac{\Delta \hat{Q}}{\langle \hat{Q}\rangle} \right|_{\rm{max}} \ll 1.
\label{relQmaxsemi}
\end{equation}
Therefore, the semiclassicality imposed on $\hat{P}$
guaranties the semiclassicality for $\hat{Q}$. We can see that
there is no cosmic forgetfulness if the condition (\ref{semiP}) is
satisfied.

The relation between  maximal fluctuations $\left. \frac{\Delta
\hat{P}}{\langle \hat{P}\rangle} \right|_{\rm{max}}$ and the
fluctuations at the bounce $\left. \frac{\Delta \hat{P}}{\langle
\hat{P}\rangle} \right|_{\rm{b}}$ is given by
\begin{equation}
\left(\left. \frac{\Delta \hat{P}}{\langle \hat{P}\rangle}
\right|_{\rm{max}} \right)^2=
\tanh(4\tilde{\alpha})+\frac{1}{\tilde{E}_0^2
\cosh(4\tilde{\alpha})}+
\frac{e^{4\tilde{\alpha}}}{\cosh(4\tilde{\alpha})} \left(\left.
\frac{\Delta \hat{P}}{\langle \hat{P}\rangle} \right|_{\rm{b}}
\right)^2,
\end{equation}
which leads to
\begin{equation}
\left. \frac{\Delta \hat{P}}{\langle \hat{P}\rangle} \right|_{\rm{max}} \geq
\frac{e^{2\tilde{\alpha}}}{\sqrt{\cosh(4\tilde{\alpha})}}
\left. \frac{\Delta \hat{P}}{\langle \hat{P}\rangle} \right|_{\rm{b}}.
\end{equation}
Therefore, if the maximal fluctuations of $\hat{P}$ are
constrained:
\begin{equation}
\left. \frac{\Delta \hat{P}}{\langle \hat{P}\rangle} \right|_{\rm{max}} \ll 1,
\end{equation}
then we have
\begin{equation}
\left. \frac{\Delta \hat{P}}{\langle \hat{P}\rangle}
\right|_{\rm{b}} \ll
e^{-2\tilde{\alpha}}\sqrt{\cosh(4\tilde{\alpha})} \ll 1,
\end{equation}
which proves  (owing to $\left. \frac{\Delta \hat{Q}}{\langle
\hat{Q}\rangle} \right|_{\rm{max}} \ll 1$, which implies $\alpha
\ll 1 $) that the bounce is semiclassical as well.

We have shown that condition $\left. \frac{\Delta \hat{P}}{\langle
\hat{P}\rangle} \right|_{\rm{max}} \ll1$ implies $\left.
\frac{\Delta \hat{Q}}{\langle \hat{Q}\rangle} \right|_{\rm{max}}
\ll1$. Can this implication be true also in the opposite
direction? The condition  $\left. \frac{\Delta
\hat{Q}}{\langle \hat{Q}\rangle} \right|_{\rm{max}} \ll1$ it
equivalent, due to (\ref{relQ}), to the  restriction
$\tilde{\alpha} \ll 1$. By taking this into account, the maximal
relative fluctuations of $\hat{P}$ can be expressed as follows
\begin{equation}
\left. \frac{\Delta \hat{P}}{\langle \hat{P}\rangle} \right|_{\rm{max}}  =
 \frac{1}{2} \frac{1}{\sqrt{\tilde{\alpha}} \tilde{E}_0}
+2  \frac{\sqrt{\tilde{\alpha}}}{\tilde{E}_0}+\mathcal{O}(\tilde{\alpha}^{3/2}).
\end{equation}
The first term in this expansion grows rapidly with decrease of
$\tilde{\alpha}$. Therefore, unless the value of $\tilde{E}_0$ is
not sufficiently large, the condition $\left. \frac{\Delta
\hat{P}}{\langle \hat{P}\rangle} \right|_{\rm{max}} \ll1$ is not
fulfilled. This condition is fulfilled if $\tilde{E}_0 \gg
1/\sqrt{\tilde{\alpha}} $, but this is exactly requirement of the
semiclassicality of the relative fluctuations of
$\hat{\mathbb{H}}$. Therefore,  if the condition
(\ref{RelHsemclass}) is fulfilled, we have the equivalence:
\begin{equation}
 \left( \left. \frac{\Delta \hat{P}}{\langle \hat{P}\rangle} \right|_{\rm{max}}
 \ll 1\right) \
\Longleftrightarrow \  \left( \left. \frac{\Delta \hat{Q}}{\langle
\hat{Q}\rangle} \right|_{\rm{max}} \ll 1 \right).
\end{equation}

To complete our considerations, we show that the saturated value
of $\Delta\hat{Q}\Delta \hat{P}$ can be easily expressed in terms
of maximal relative fluctuations of $\hat{Q}$ and $\hat{P}$:
\begin{equation}
\lim_{s\rightarrow \pm \infty} \Delta\hat{Q}\Delta \hat{P} =
\frac{1}{2} e^{4\tilde{\alpha}} \tilde{E}_0 \left. \frac{\Delta
\hat{Q}}{\langle \hat{Q}\rangle} \right|_{\rm{max}} \left.
\frac{\Delta \hat{P}}{\langle \hat{P}\rangle}
\right|_{\rm{max}} \geq \frac{1}{2}.
\end{equation}
Owing to Heisenberg's uncertainty relation, the above expression
leads to the following constraint
\begin{equation}
\left. \frac{\Delta
\hat{Q}}{\langle \hat{Q}\rangle} \right|_{\rm{max}} \left.
\frac{\Delta \hat{P}}{\langle \hat{P}\rangle}
\right|_{\rm{max}} \geq  \frac{e^{-4\tilde{\alpha}}}{\tilde{E}_0}.
\end{equation}
This constraint, together with inequality (\ref{PmaxQmax}),
results gives
\begin{equation}
\left.\frac{\Delta \hat{P}}{\langle \hat{P}\rangle}
\right|_{\rm{max}} \geq  \frac{e^{-2\tilde{\alpha}}}{\sqrt{\tilde{E}_0}}.
\end{equation}
Thus, the samiclassicality condition (\ref{semiP}) leads to the
following constraint:
\begin{equation}
\tilde{E}_0 \gg e^{-2\tilde{\alpha}}.
\end{equation}

\subsection{Forgetfulness}

Let us try to answer the question: Was the Universe quantum
or semiclassical {\it before} the big bounce? This question is
related to the problem of cosmic forgetfulness discussed recently
in papers
\cite{BojowaldNature,Corichi:2007am,Bojowald:2007gc,Corichi:2011rt,Kaminski:2010yz}.
If the Universe was {\it quantum} before the big bounce and the
bounce transition  turned it into {\it semiclassical} one, we can
talk about a sort of cosmic {\it amnesia}. In this case the
complete information about the Universe {\it before} the  bounce
cannot be obtained from the observational data {\it after} the
bounce.

The above question  has been addressed so far mainly by an
examination of the relative fluctuations of the volume observable
$\hat{v}$ (proportional to our $\hat{P}$ observable)
\cite{Corichi:2011rt,Kaminski:2010yz}. However, as we have shown,
this type of fluctuation is {\it symmetric} with respect to the
bounce. Therefore, the constraint on relative fluctuation at some
time $+|s|$ results with the same constraint on the fluctuations
at the time $-|s|$. As one can see on Fig. \ref{Fig6},  the
fluctuations of $\hat{P}$ saturate quickly  outside the
neighborhood of the bounce (within a few Planck's times). Thus,
present cosmic observation  of the semiclassicality, $\frac{\Delta
\hat{P}}{\langle \hat{P}\rangle}= \frac{\Delta \hat{v}}{\langle
\hat{v}\rangle} \ll 1$,  of the Universe would guarantee its
semiclassicality in the distant past before the bounce. However,
the situation is that this type of relative fluctuation cannot be
`measured'. The reason is that one does not know, first of all,
how to measure  the ${\langle \hat{v}\rangle}$ quantity. It could
be possible to measure the volume, to some extent, if the Universe
was curved and the curvature term was measured in astronomical
observations. At present, there is however no indication for such
a contribution. Therefore, the fluctuations of $\hat{v}$ are not
measurable so the cosmic forgetfulness cannot be examined by using
the relative fluctuations $\frac{\Delta \hat{v}}{\langle
\hat{v}\rangle}$.

What about the $\hat{Q}$ observable?  The variable $Q$ is directly
related to the expansion rate, i.e. the Hubble factor
(\ref{Hubbledef}), which is a quantity that can be determined
observationally. Thus, the value of $\langle \hat{Q}\rangle$ can
be measured. Also the observational uncertainty of the Hubble
factor can be used to constrain $ \Delta \hat{Q}$. Therefore,
relative fluctuations of $\hat{Q}$ can be constrained
observationally! The present value of the Hubble factor is
$H_0=70.2 \pm 1.4 \ \rm{km} \ \rm{s}^{-1}\
\rm{Mpc}^{-1}$ \cite{Komatsu:2010fb}, therefore $\frac{\sigma(H_0)}{H_0} \approx 0.02$.
As we have shown in Sec. II,  in the classical limit $Q=\gamma H$. Thus, we
propose to consider the constraint:  $\frac{\Delta \hat{Q}}{\langle \hat{Q}\rangle} < 0.02$.
This is because the relative quantum fluctuations cannot be
greater than the relative uncertainty of measurement.

Due to the relation (\ref{trel}),  the directions of the intrinsic
time $T$ and the coordinate time $t$ are the opposite. Therefore,
relative fluctuations of $\hat{Q}$ at $T\rightarrow -\infty$
correspond to the limit $t\rightarrow +\infty$. Thus, the relative
fluctuations of $\hat{Q}$ grow in the coordinate time $t$ and
saturate at $\sqrt{e^{4\tilde{\alpha}}-1}$.  The model we consider
is applicable to the evolution in vicinity of the Planck epoch.
However, if we assume that the relative fluctuations  behave
similarly threafter, the present restriction $\frac{\Delta
\hat{Q}}{\langle \hat{Q}\rangle} < 0.02$ can be used to constraint
the model. The condition $\sqrt{e^{4\tilde{\alpha}}-1}< 0.02$
translates into the condition $\tilde{\alpha} < 10^{-4}$.
If the Gaussian state with such value of its parameter can be
treated as a semiclassical one, we can say that the amnesia does
not occur. In such a case the contraction and the bounce phases
are semiclassical, so we have the cosmic {\it recall}.

The above constrains are quite preliminary. There exist the
possibility to put more robust  constraints based on the phase of
inflation and observations of the Cosmic Microwave Background
Radiation. However, for this purpose the model has to be
generalized by taking into account potential of the scalar field.

\section{Summary and Conclusions}

In the Hamiltonian formulation of general relativity, GR, the
total Hamiltonian is a linear combination of the constraints (see,
e.g. Eq. (\ref{conham})) so it cannot play the role of the
generator of an evolution of a gravitational system. On the other
hand, the GR system evolves according to the Einstein equations.
Can one overcome this difficulty of the Hamiltonian
formulation? There are two ways of dealing with this problem:
\begin{enumerate}
    \item One eliminates the time variable in favor of any canonical
     variable by some formal trick carried out on Hamilton's equations (see,
     e.g. \cite{Dzierzak:2009ip} for more details), which leads to the  so
     called {\it relative} dynamics commonly used in LQC. In this
     procedure the constraints are used to define the {\it physical} phase space.
     However, the evolution is poorly defined since one gets the dependance
     of canonical variables in terms of any specific variable so
     the dependance on time may become deeply hidden.
    \item Using the constraints, one expresses the specific canonical variable
    (of Hamilton's equations) in terms of other variables. Elementary Dirac
    observables are constants of motion and include the dynamical constraint.
    The new Hamilton's equations (including the constraint) are used for
    finding the Hamiltonian that generates dynamics on the physical
    phase space. This new Hamiltonian is no longer a dynamical constraint of the
    classical theory, but the generator of dynamics, formally {\it free}  of dynamical
    constraints. This approach  restores  the notion of an evolution of a classical system.
\end{enumerate}

One should remember that the above considerations on the
evolution parameter apply first of all to the situation with
`matter' field, like the scalar field, that can be used to define
this parameter. It would be interesting to extend these ideas to
the case with no sources. The simplest  nontrivial example is the
Kasner model.

When we wish to {\it quantize} a Hamiltonian system with
constraints, we may apply the Dirac or the RPS quantization
methods. Both methods are  plagued by numerous {\it ambiguities}.
Our analyzes are devoid of the need of any restriction
to some superselection sectors that naturally arise in the
standard LQC quantization.
The latter need not be a drawback of the method as it may leave
some interesting imprints in the physical predictions  \cite{GJT}.
The former one, seems to be less complicated than the Dirac
method, and offers an {\it analytical} insight into physical
aspect of considered model. Presented results demonstrate, to some
extent, that our quantization scheme leads to a quantum {\it
cosmological} system with general properties of a quantum systems
we are dealing with in {\it terrestrial} laboratories.

It seems that one can apply our method to much more complex
cosmological models than the FRW universe. Recently, we have
managed to quantize the Bianchi I model
\cite{Dzierzak:2009dj,Malkiewicz:2010py}.  The case of the Biachi
II model can be treated by analogy. In summary, we suggest that
quantization of cosmological systems in terms of the RPS method is
much more efficient and unique than Dirac's method. However, since
quantum cosmology `experimental' data are not available yet, the
best strategy seems to be applying both methods to {\it compare}
the results. An agreement of the results would prove that the
procedure of quantization was correct.

The loop quantization methods applied to simple cosmological
models teach us that approximating the curvature of connection by
holonomies around small loops enable replacing   classical
sinularities by quantum bounces
\cite{Ashtekar:2003hd,Bojowald:2006da,Ashtekar:2006wn,
Dzierzak:2008dy,Dzierzak:2009ip,Malkiewicz:2009qv,Malkiewicz:2009zd,
Mielczarek:2010rq,Mielczarek:2010wu,Dzierzak:2009dj,Malkiewicz:2010py}.
To get some information about the nature  of a bounce, one
examines propagation   quantum states across the bounce
\cite{Corichi:2007am,BojowaldNature,Bojowald:2007gc,Corichi:2011rt,
Kaminski:2010yz}. Such method is similar, to some extent, to the
method used, for instance, in nuclear physics where one scatters a
particle against an atomic nucleus to get information on the
structure of the latter. In papers
\cite{BojowaldNature,Bojowald:2007gc} one considers solvable toy
models (motivated by LQC) to argue that a quantum state before the
bounce may become decohered at the bounce and become semiclassical
afterwards. Applying the sLQC prescription authors of
\cite{Corichi:2007am,Corichi:2011rt} claim that there is no
cosmological amnesia at all: suitable semiclassical state before
the bounce keeps being semiclassical after the bounce.  Authors of
\cite{Kaminski:2010yz} give strong support to this result by
applying various analytical and numerical  methods, within the
standard LQC, to general forms of semiclassical states. They
identify the condition under which one has the preservation of the
semiclassicality across the bounce. However, they have mainly
examined  the volume observable  which is little useful for
testing the cosmic amnesia.

In our paper we have considered the transition of the Gaussian
wave packet across the bounce of the quantum FRW universe within
an {\it exact} framework.   It results from our studies that the
$\hat{Q}$ observable is the proper quantity to study the cosmic
amnesia. It is because relative fluctuations of $\hat{Q}$ can be
observationally constrained. Moreover, the semiclassicality
condition imposed in the expanding phase restricts also the
quantum fluctuations in the contracting phase.  The preliminary
observational constraint $\;\frac{\Delta \hat{Q}}{\langle
\hat{Q}\rangle} < 0.02\;$ indicates  that the semiclassicality
condition,  as defined earlier, may be fulfilled. Owing to this,
one can infer that there was a cosmic {\it amnesia} or there was a
{\it recall} depending on what we mean by a semiclassical state.
Our results support the prediction of the standard LQC
\cite{Corichi:2007am,Corichi:2011rt,Kaminski:2010yz}. We suggest
to repeat the calculations with the  variety of  states different
from the Gaussian type  states to verify  our results.

On the other hand, the LQC results obtained for the FRW type
models cannot be probably used successfully to describe the
Universe. The very high symmetry of space specific to the FRW
model is probably unrealistic near the cosmological singularity.
We suggest that the real nature of the bounce may become known
only after we quantize the Belinskii-Khalatnikov-Lifshitz (BKL)
scenario \cite{Belinski:2009wj,BKL2,BKL3}, which concerns the
generic cosmological singularity. Quantization of simple
cosmological models carried out during the last decade may be
treated as warming up before meeting this challenge.

\ack
We thank Vladimir Belinski, Jean-Pierre Gazeau,
Przemys{\l}aw Ma{\l}kiewicz and Wies{\l}aw Pusz  for helpful
discussions. JM has been supported by Polish Ministry of Science
and Higher Education grant N N203 386437 and by Foundation of
Polish Science scholarship START. Also we would like to thank the
anonymous referees for the constructive criticisms.

\appendix

\section{Quantum asymptotics of $\hat{Q}$.}

It this appendix we study dispersions and relative fluctuations of
observable $\hat{O}$ in the limits $s \rightarrow \pm \infty$. We
show these limits  can be found analytically by performing
suitable expansions of integrals  in expressions (\ref{expectQ})
and (\ref{expectQ2}). Based on this, we derive equations
(\ref{relQpinf}) and (\ref{relQminf}).

\subsection{The case $s \rightarrow + \infty$ }

Let us introduce $\epsilon := e^{-2\frac{s}{t_{\rm{Pl}}}}$, which tends
to zero for $s \rightarrow + \infty$.
Based on this, one can perform Taylor expansion with respect to $\epsilon$, as follows
\begin{equation}
\arctan\left\{ \frac{e^{2y}}{\epsilon}\right\} = \frac{\pi}{2}-e^{-2y}
\epsilon +\mathcal{O}(\epsilon^3).
\label{Taylor1}
\end{equation}
This expansion applied to equation  (\ref{expectQ}) gives
\begin{eqnarray}
{\langle \hat{Q} \rangle} &=& \frac{1}{\sqrt{2\pi \tilde{\alpha}}} \frac{2}{\lambda}
\int_{-\infty}^{+\infty}   \arctan\left\{ \frac{e^{2y}}{\epsilon} \right\} \exp \left\{-
\frac{1}{2\tilde{\alpha}}y^2 \right\} dy \nonumber  \\
&=& \frac{\pi}{\lambda} - \frac{2}{\lambda}e^{2\tilde{\alpha}} \epsilon +\mathcal{O}(\epsilon^3).
\label{expQpinf}
\end{eqnarray}
By squaring expansion (\ref{Taylor1}), we obtain
\begin{equation}
\arctan^2\left\{ \frac{e^{2y}}{\epsilon}\right\} = \left(  \frac{\pi}{2}\right)^2 -
\pi e^{-2y} \epsilon + e^{-4y}\epsilon^2+\mathcal{O}(\epsilon^3).
\end{equation}
This expansion, applied in equation (\ref{expectQ2}), leads to
\begin{eqnarray}
\langle \hat{Q}^2 \rangle &=& \frac{1}{\sqrt{2\pi
\tilde{\alpha}}} \left(\frac{2}{\lambda}\right)^2
\int_{-\infty}^{+\infty}   \arctan^2\left\{    \frac{e^{2y}}{\epsilon} \right\} \exp \left\{-
\frac{1}{2\tilde{\alpha}}y^2 \right\} dy \nonumber \\
&=&   \left( \frac{\pi}{\lambda}\right)^2 -
\pi  \left( \frac{2}{\lambda}\right)^2 e^{2\tilde{\alpha}} \epsilon+
\left(\frac{2}{\lambda}\right)^2 e^{4\tilde{\alpha}}\epsilon^2 +\mathcal{O}(\epsilon^3).
\label{expQ2pinf}
\end{eqnarray}
Based on (\ref{expQpinf}) and (\ref{expQ2pinf}), we find
\begin{equation}
\langle \hat{Q}^2 \rangle - ({\langle \hat{Q} \rangle})^2 =
\left(\frac{2}{\lambda}\right)^2e^{4\tilde{\alpha}}\left(e^{4\tilde{\alpha}}-1\right)\epsilon^2+
\mathcal{O}(\epsilon^4),
\end{equation}
which leads to the following expression for the dispersion of
$\hat{Q}$:
\begin{equation}
\Delta \hat{Q} =\sqrt{ \langle \hat{Q}^2 \rangle - ({\langle \hat{Q} \rangle})^2 }=
\epsilon \frac{2}{\lambda}e^{2\tilde{\alpha}} \sqrt{e^{4\tilde{\alpha}}-1}
+\mathcal{O}(\epsilon^3).
\label{DeltaQpinf}
\end{equation}
Using  (\ref{expQpinf}) and (\ref{DeltaQpinf}) we get
\begin{equation}
\lim_{\epsilon \rightarrow 0} \frac{\Delta \hat{Q}}{{\langle \hat{Q} \rangle}} =
\lim_{\epsilon \rightarrow 0}
\left[\epsilon \frac{2}{\pi}e^{2\tilde{\alpha}} \sqrt{e^{4\tilde{\alpha}}-1}
+\mathcal{O}(\epsilon^3)\right] =0,
\end{equation}
which proofs equation (\ref{relQpinf}).

\subsection{The case $s \rightarrow - \infty$ }

Let us introduce the variable $\epsilon :=
e^{2\frac{s}{t_{\rm{Pl}}}}$, which tends to zero for $s
\rightarrow - \infty$. It is worth to stress that the parameter
$\epsilon$ introduced here differs from that used in the previous
section. We perform the Taylor expansion
\begin{equation}
\arctan\left\{ e^{2y}\epsilon \right\} = e^{2y} \epsilon +\mathcal{O}(\epsilon^3).
\label{Taylor2}
\end{equation}
This expansion, applied in equation (\ref{expectQ2}), leads to
\begin{eqnarray}
{\langle \hat{Q} \rangle} &=& \frac{1}{\sqrt{2\pi \tilde{\alpha}}} \frac{2}{\lambda}
\int_{-\infty}^{+\infty}   \arctan\left\{ e^{2y}\epsilon \right\} \exp \left\{-
\frac{1}{2\tilde{\alpha}}y^2 \right\} dy \nonumber  \\
&=& \frac{2}{\lambda}e^{2\tilde{\alpha}} \epsilon +\mathcal{O}(\epsilon^3).
\label{expQminf}
\end{eqnarray}
By squaring (\ref{Taylor2}), and applying it to equation (\ref{expectQ2}), we find
\begin{eqnarray}
\langle \hat{Q}^2 \rangle &=& \frac{1}{\sqrt{2\pi
\tilde{\alpha}}} \left(\frac{2}{\lambda}\right)^2
\int_{-\infty}^{+\infty}   \arctan^2\left\{    e^{2y}\epsilon \right\} \exp \left\{-
\frac{1}{2\tilde{\alpha}}y^2 \right\} dy \nonumber \\
&=& \left(\frac{2}{\lambda}\right)^2 e^{8\tilde{\alpha}}\epsilon^2 +\mathcal{O}(\epsilon^4).
\label{expQ2minf}
\end{eqnarray}
Expansions  (\ref{expQminf}) and (\ref{expQ2minf}) lead to
\begin{equation}
\langle \hat{Q}^2 \rangle - ({\langle \hat{Q} \rangle})^2 =
\left(\frac{2}{\lambda}\right)^2e^{4\tilde{\alpha}}\left(e^{4\tilde{\alpha}}-1\right)\epsilon^2+
\mathcal{O}(\epsilon^4).
\end{equation}
Based on this, dispersion of $\hat{Q} $ in the limit $s \rightarrow - \infty$  is given by
\begin{equation}
\Delta \hat{Q} =\sqrt{ \langle \hat{Q}^2 \rangle - ({\langle \hat{Q} \rangle})^2 }=
\epsilon \frac{2}{\lambda}e^{2\tilde{\alpha}} \sqrt{e^{4\tilde{\alpha}}-1}
+\mathcal{O}(\epsilon^3).
\label{DeltaQminf}
\end{equation}
With use of  (\ref{expQminf}) and (\ref{DeltaQminf}) we find that
\begin{eqnarray}
\lim_{\epsilon \rightarrow 0} \frac{\Delta \hat{Q}}{{\langle \hat{Q} \rangle}} =
\lim_{\epsilon \rightarrow 0}
\frac{ \epsilon \frac{2}{\lambda}e^{2\tilde{\alpha}} \sqrt{e^{4\tilde{\alpha}}-1}
+\mathcal{O}(\epsilon^2)}{ \frac{2}{\lambda}e^{2\tilde{\alpha}} \epsilon +\mathcal{O}(\epsilon^3)}
=\lim_{\epsilon \rightarrow 0} \left[ \sqrt{e^{4\tilde{\alpha}}-1}+\mathcal{O}(\epsilon^2) \right] \\
=\sqrt{e^{4\tilde{\alpha}}-1},
\end{eqnarray}
which proofs equation (\ref{relQminf}).

\end{document}